\documentclass[prb, aps, twocolumn, floatfix, superscriptaddress]{revtex4-2}
\usepackage{amsmath}
\usepackage{amssymb}
\usepackage{graphicx}
\usepackage{color}

\newcommand{\be}{\begin{equation}}
\newcommand{\ee}{\end{equation}}
\newcommand{\bea}{\begin{eqnarray}}
\newcommand{\eea}{\end{eqnarray}}
\newcommand{\bse}{\begin{subequations}}
\newcommand{\ese}{\end{subequations}}
\newcommand{\emb}{${\rm EuMg_2Bi_2}$}
\newcommand{\ems}{${\rm EuMg_2Sb_2}$}
\newcommand{\cas}{${\rm CaAl_2Si_2}$}
\newcommand{\esa}{${\rm EuSn_2As_2}$}
\newcommand{\eag}{${\rm EuAl_2Ge_2}$}
\newcommand{\eca}{${\rm EuCd_2As_2}$}
\newcommand{\eia}{${\rm EuIn_2As_2}$}
\newcommand{\eg}{${\rm EuGa_4}$}

\begin{document}

\title{Anisotropic magnetism and electronic structure of trigonal EuAl$_2$Ge$_2$ single crystals}

\author{Santanu Pakhira}
\affiliation{Ames National Laboratory, Iowa State University, Ames, Iowa 50011, USA}
\author{Asish K. Kundu}
\affiliation{Condensed Matter Physics and Materials Science Division, Brookhaven National Laboratory, Upton, New York 11973, USA}
\author{Farhan Islam}
\affiliation{Ames National Laboratory, Iowa State University, Ames, Iowa 50011, USA}
\affiliation{Department of Physics and Astronomy, Iowa State University, Ames, Iowa 50011, USA}
\author{M. A. Tanatar}
\affiliation{Ames National Laboratory, Iowa State University, Ames, Iowa 50011, USA}
\affiliation{Department of Physics and Astronomy, Iowa State University, Ames, Iowa 50011, USA}
\author{Tufan Roy}
\affiliation{Center for Science and Innovation in Spintronics, Core Research Cluster,  Tohoku University, Sendai 980-8577, Japan}
\author{Thomas Heitmann}
\affiliation{The Missouri Research Reactor and Department of Physics and Astronomy, University of Missouri, Columbia, Missouri 65211, USA}
\author{T. Yilmaz}
\affiliation{National Synchrotron Light Source II, Brookhaven National Laboratory, Upton, New York 11973, USA}
\author{E. Vescovo}
\affiliation{National Synchrotron Light Source II, Brookhaven National Laboratory, Upton, New York 11973, USA}
\author{Masahito Tsujikawa}
\affiliation{Research Institute of Electrical Communication, Tohoku University, Sendai 980-8577, Japan}
\author{Masafumi Shirai}
\affiliation{Center for Science and Innovation in Spintronics, Core Research Cluster, Tohoku University, Sendai 980-8577, Japan}
\affiliation{Research Institute of Electrical Communication, Tohoku University, Sendai 980-8577, Japan}
\author{R. Prozorov}
\affiliation{Ames National Laboratory, Iowa State University, Ames, Iowa 50011, USA}
\affiliation{Department of Physics and Astronomy, Iowa State University, Ames, Iowa 50011, USA}
\author{David Vaknin}
\affiliation{Ames National Laboratory, Iowa State University, Ames, Iowa 50011, USA}
\affiliation{Department of Physics and Astronomy, Iowa State University, Ames, Iowa 50011, USA}
\author{D. C. Johnston}
\affiliation{Ames National Laboratory, Iowa State University, Ames, Iowa 50011, USA}
\affiliation{Department of Physics and Astronomy, Iowa State University, Ames, Iowa 50011, USA}

\date{\today}

\begin{abstract}
Understanding the interplay between magnetic and electronic degrees of freedom is of profound recent interest in different Eu-based magnetic topological materials. In this work, we studied the magnetic and electronic properties of the layered Zintl-phase compound \eag\ crystallizing in the trigonal \cas-type structure. We report zero-field neutron diffraction, temperature~$T$- and magnetic-field~$H$-dependent magnetic susceptibility $\chi(T, H)$, isothermal magnetization $M(T, H)$, heat capacity $C_{\rm p}(T, H)$, and electrical resistivity $\rho(T, H)$ measurements, together with \mbox{$T$-dependent} angle-resolved photoemission spectroscopy (ARPES) \mbox{measurements} complemented with first-principle calculations. \eag\ undergoes second-order \mbox{A-type} antiferromagnetic (AFM) ordering below $T_{\rm N} = 27.5(5)$~K, with the Eu moments (Eu$^{2+},\, S= 7/2$) aligned ferromagnetically in the $ab$~plane while these layers are stacked antiferromagnetically along the $c$~axis. The critical fields at which all moments become parallel to the field are 37.5(5) and 52.5(5)~kOe for $H\parallel ab$ and $H\parallel c$, respectively.  The $H = 0$ magnetic structure consists of trigonal AFM domains associated with \mbox{$ab$-plane} magnetic anisotropy and a field-induced reorientation of the Eu spins in the domains is also evident at $T = 2$~K below the critical field $H_{c1} = 2.5(1)$~kOe. The $\rho(T)$ measurements reveal metallic behavior transforming into a slight resistivity increase on cooling towards $T_{\rm N}$\@.  A pronounced loss of spin-disorder scattering is observed below $T_{\rm N}$. The ARPES results show that \eag\ is metallic both above and below $T_{\rm N}$, and the Fermi surface is anisotropic with two hole pockets at the zone center and one small electron pocket at each M~point.  In the AFM phase, we directly observe folded bands in ARPES due to the doubling of the magnetic unit cell along the $c$~axis with an enhancement of quasiparticle weight due to the complex change in the coupling between the magnetic moments and itinerant electrons on cooling below  $T_{\rm N}$. The observed electronic structure is well reproduced by first-principle calculations, which also predict the presence of nontrivial electronic states near the Fermi level in the AFM phase with $Z_2$ topological numbers~1;(000).

\end{abstract}

\maketitle

\section{Introduction}

It is rewarding to study different classes of novel quantum materials having a complex interplay of lattice, spin, and electronic degrees of freedom. These materials can exhibit a plethora of interesting physical properties including superconductivity, heavy fermion behavior, quantum phase transitions, complex magnetic order, magnetic frustration, valence fluctuations, and nontrivial topological phases. One such family of materials is comprised of Zintl-phase compounds that have gained significant recent interest owing to the complex interplay of magnetic and electronic degrees of freedom. These materials exhibit topological states, proximity between metal-semimetal-semiconductor-insulator phases, anomalous and topological Hall effects, low-field-induced spin reorientations within  antiferromagnetic (AFM) domains, along with large thermoelectricity as recently reported in various \mbox{compounds~\cite{Zheng1986, Shuai2017, Wang2018, Riberolles2021, Zhu2016, Ogunbunmi2021, Varnava2022, Yan2022, Kundu2022a, Pakhira2022a}}.

Many $AM_2X_2$-type Zintl-phase compounds have been investigated, where $A$ is an alkaline or lanthanide element, $M$ is a metallic  $sp$ element, and $X$ is an $sp$-element anion where the $A$ atom has either a planar triangular or square-lattice structure.  These materials have recently been reported to exhibit electronic states having nontrivial band topology.  These states include a topological insulating state, a Dirac/Weyl-type semimetallic state, or an axion-insulating state, and are attractive candidates for dissipationless electron transport~\cite{Kundu2022a, Li2019, Jo2020, Xu2019, Rahn2018, Kundu2022b, kabir2019, Marshall2020}. It has been experimentally found that when the $A$~site of these compounds is fully or partially occupied by a rare-earth element, the materials show enhanced carrier mobility and carrier concentration compared to those with $A$ as an alkaline-earth metal~\cite{May2012, Toberer2010}; the origin of this behavior is currently unknown.

For example, such magnetic Eu-based compounds are of significant interest due to their complex interplay of magnetism and band topology, as reported for \eia, \eca, \emb, and \esa~\cite{Riberolles2021, Yan2022, Li2019, Jo2020, Xu2019, Rahn2018, Marshall2020, Pakhira2020, Pakhira2021, Marshall2022, Pakhira2021a, Lv2022}.  The magnetic properites associated with different anisotropy energies could thus also play an important role in tuning the electronic states in these materials associated with magnetic ordering.  Although the Eu$^{2+}$ ion with spin $S=7/2$ and orbital angular momentum $L = 0$ exhibits negligible single-ion anisotropy, the magnetic properties in most of these materials are anisotropic~\cite{Pakhira2020, Pakhira2021, Pakhira2021a, ZCWang2022, Berry2022, Berry2021, Pakhira2022}. Here the anisotropy arises from magnetic-dipole and/or anisotropic RKKY interactions.

To further investigate the properties of this class of materials,  here we report the growth of \eag\ single crystals with the trigonal CaAl$_2$Si$_2$ crystal structure~\cite{Wartenberg2002} and studies of their magnetic, electronic-transport, and electronic-structure properties.  These include zero-field neutron-diffraction measurements of the ordered magnetic structure, temperature $T$- and magnetic-field~$H$-dependent magnetization~$M(H,T)$, heat capacity~$C_{\rm p}(T)$, and electrical-resistivity~$\rho(H,T)$ measurements, along with $T$-dependent angle-resolved photoemission spectroscopy (ARPES) studies of the electronic structure. The experimental electronic structure is mapped by calculating the band structure of \eag\ using density-functional theory (DFT).

We find that \eag\ is metallic as revealed by the $\rho(T)$ and ARPES measurements complemented with theoretical band-structure calculations. The neutron-diffraction experiments demonstrate that \eag\ exhibits A-type AFM order below its N\'eel temperature $T_{\rm N} = 27.5(5)$~K\@.  In this magnetic structure the Eu$^{2+}$ moments $\mu = 7~\mu_{\rm B}$ with spectroscopic-splitting factor $g=2$ and spin $S=7/2$ are aligned ferromagnetically in each $ab$-plane layer, where the moments in adjacent layers along the $c$~axis are aligned antiferromagnetically. The $C_{\rm p}$ data for $H=0$ exhibit a second-order $\lambda$-type peak at $T_{\rm N}$. The ARPES results further reveal magnetism-induced band folding and enhancement of the quasiparticle intensity in the magnetic ground state. Splitting of the energy bands is observed even above $T_{\rm N}$, possibly related to time-reversal-symmetry breaking associated with short-range ferromagnetic (FM) correlations above $T_{\rm N}$.

Over the broad field range $0\leq H\leq 55$~kOe, the $M(H)$ data at $T= 2$~K appear to be linear for both $H\parallel ab$ and $H\parallel c$ with respective  critical fields  $H^{\rm c}_{ab} = 37.5$ and $H^{\rm c}_{c} = 52.5$~kOe, at which all moments become parallel to the respective field.  However, detailed  \mbox{$M(H_{ab},T=2$~K)}  measurements at low fields $H_{ab}\leq H_{\rm c1} = 2.5$~kOe exhibit anomalous positive curvature, whereas for $H_{ab}> H_{\rm c1}$ the data are again linear up to $H^{\rm c}_{ab}$.   This behavior is quantitatively described by a model where the A-type AFM structure occurs in three  trigonal domains in which the Eu moments in each domain are aligned at 120$^\circ$ to each other in $H=0$.  With increasing $H_{ab}$ the moments in each domain reorient to become perpendicular to ${\bf H}_{ab}$ until $H_{\rm c1}$ is reached, above with all moments progressively cant towards ${\bf H}_{ab}$ until  $H^{\rm c}_{ab}$ is attained.

Experimental and theoretical details are given in Sec.~\ref{Sec:ExpDet}.  The results and discussion of the various measurements and analyses are presented in Sec.~\ref{Sec:Results}, and concluding remarks are provided in Sec.~\ref{Sec:Conclu}.

\section{\label{Sec:ExpDet} Experimental and Theoretical Details}

Single crystals of \eag\ were grown using the flux method with starting composition Eu:Al:Ge = 1:20:2. The Eu (Ames Lab), Al (Alfa Aesar, 99.9995\%), and Ge (Alfa Aesar, 99.9999\%) were loaded into a 2 mL alumina crucible and sealed in a silica tube under $1/4$ atm high-purity argon. The assembly was heated to 1175~$^{\circ}$C inside a box furnace at a rate of 100~$^{\circ}$C/h. After holding the temperature for 6~h, the furnace was cooled to 700~$^{\circ}$C at a rate of 10~$^{\circ}$C/h. The assembly was then centrifuged to separate the crystals from the molten flux. Hexagonal plate-like crystals with typical dimensions $3\times3\times2$ mm$^3$ were obtained from the growth. The homogeneity and chemical composition of the crystals were confirmed using a JEOL scanning-electron microscope (SEM) equipped with an energy-dispersive x-ray spectroscopy (EDS) analyzer.  The magnetic measurements were carried out using a Magnetic-Properties-Measurement System (MPMS) from Quantum Design, Inc., in the $T$ range 1.8--300 K and with $H$ up to 5.5 T (1~T~$\equiv10^4$~Oe).

A Physical Properties Measurement System (PPMS, Quantum Design, Inc.) was used to measure $C_{\rm p}(T)$ and $\rho(T)$ in the $T$ range 1.8--300~K and $H$ up to 9~T\@.  Four-probe $\rho(T)$ measurements were performed.  The measurements were performed on as-grown single crystals. Due to the sensitivity of EuAl$_2$Ge$_2$ to the ambient environment leading to a rapid sample decomposition, the crystals were not shaped into resistivity bars with precision geometric-factor control by polishing and cutting. However, the crystals had natural shapes suitable for in-plane resistivity measurements, having a length at least 3 times larger than the width and thickness. Resistivity measurements were performed along arbitrary directions in the $ab$~plane.  In all resistivity measurements the magnetic field was oriented transverse to the current direction.  Contacts to the fresh surfaces of the crystals were made by attaching 50~$\mu$m-diameter silver wires with In solder and mechanically reinforcing the contact with DuPont 4929N silver paint~\cite{Tanatar2016}. The contact resistance was typically in the $\Omega$  range. After application of the contacts was complete, the samples were covered with Apiezon N-grease to provide temporal protection from degradation.  For measurements in magnetic fields oriented along the $c$~axis and $ab$~plane, the samples were attached with Apiezon N-grease to the sides of a plastic cube. This provides alignment with about $\pm5^{\circ}$ accuracy~\cite{Kgoodcrystals}.

Single-crystal neutron-diffraction experiments were performed in $H=0$ using the TRIAX triple-axis spectrometer at the University of Missouri Research Reactor (MURR). An incident neutron beam of energy  30.5 meV was directed at the sample using a pyrolytic graphite (PG) monochromator. A PG analyzer was used to reduce the background. Neutron wavelength harmonics were removed from the beam using PG filters placed before the monochromator and in between the sample and analyzer. Beam divergence was limited using collimators before the monochromator; between the monochromator and sample; sample and analyzer; and analyzer and detector of $60^\prime-60^\prime-40^\prime-40^\prime$, respectively. A $\approx20$~mg \eag\ crystal was mounted on the cold tip of an Advanced Research Systems closed-cycle refrigerator with a base temperature of 4~K\@. The crystal was aligned in the $(HHL)$  scattering planes. The lattcie parrameters at base temperature are $a=4.19(1)$ and $c = 7.27(1)$ {\AA}.

ARPES experiments were performed at the Electron Spectro Microscopy (ESM) 21-ID-1 beamline of the National Synchrotron Light Source II, USA. The beamline is equipped with a Scienta DA30 electron analyzer, with base pressure better than $\sim$ 1$\times$10$^{-11}$ mbar.  Prior to the ARPES experiments, samples were cleaved inside an ultra-high vacuum chamber (UHV) at $\sim$ 9 K. All the measurements were performed using horizontally polarized light. The uncertainty in the temperature values for the ARPES measurements is $\pm 2$~K\@.

The Vienna {\it ab initio} simulation package was used for the first-principles calculations~\cite{Kresse1996, Kresse1999}. For the exchange and correlation energy/potential we used the PBE functional~\cite{Perdew1996}. The projected-augmented-wave~\cite{Bloch1994} method was used to represent the core electrons. The cut-off energy for the plane waves was set to 550 eV\@. A $k$-mesh of 14$\times$14$\times$7 (AFM phase) and 14$\times$14$\times$12 (PM phase) was used for the Brillouin-Zone integration. Spin-orbit-coupling (SOC) was considered in all calculations. The GGA $+$ U method~\cite{Liechtenstein1995} was used to treat the correlation effects of 4$f$ states in Eu. Furthermore, WANNIER90 was used for the construction of the first-principle tight binding Hamiltonian and constant energy surfaces~\cite{Mostofi2014}. The $s$ and $p$ orbitals of Ge and Al and $s$, $p$, $d$, and $f$ orbitals of Eu were used to construct maximally-locallized Wannier functions. In the case of the PM phase, we treated the $f$ electrons of Eu as core states.  The {\mbox WANNIERTOOLS} package was used for the calculation of the $Z_2$ topological number~\cite{Wu2018}.  For the visualization of the Fermi surfaces, we used FermiSurfer~\cite{Kawamura2019}

\begin{figure*}
\centering
\includegraphics[width=\linewidth]{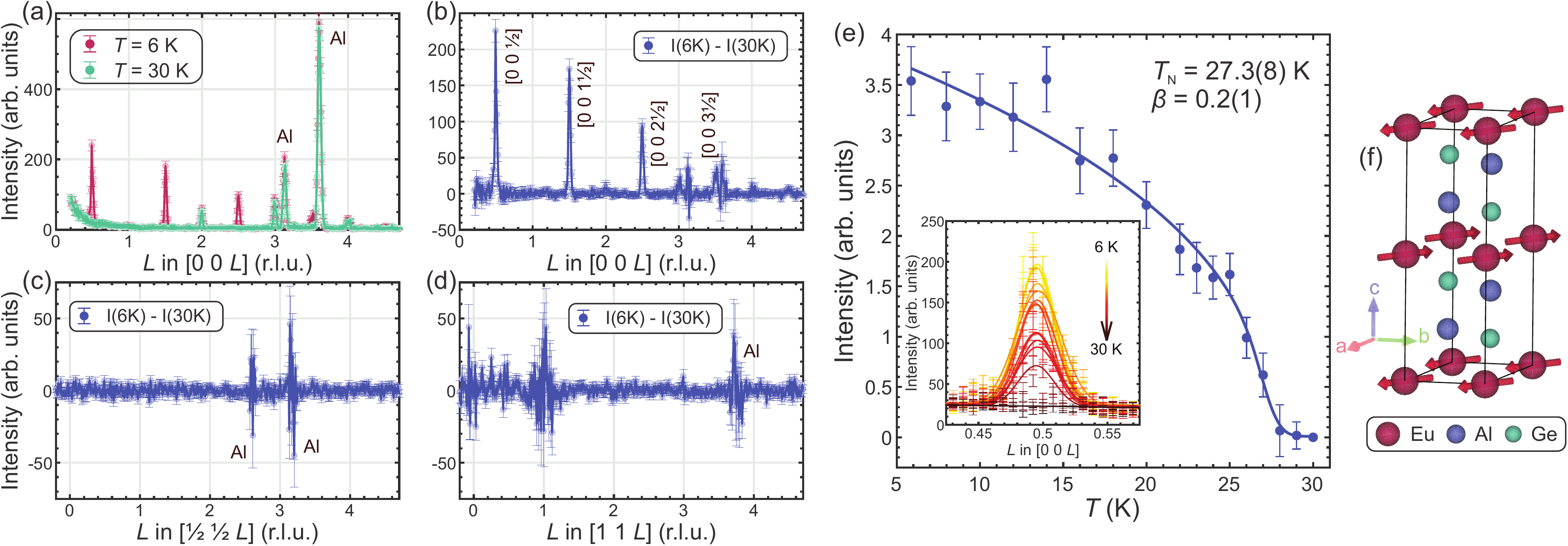}
\caption{(a) Zero-field neutron-diffraction pattern along $(00L)$ of single-crystal \eag~at 6 and 30~K, as indicated. The aluminum Bragg reflections marked on the figure originate from the sample holder. The magnetic Bragg reflections are obtained by subtracting the diffraction pattern at 30~K from the one at 6~K for (b)~$(0 0 L)$, (c)~$(\frac{1}{2} \frac{1}{2} L)$, and (d)~$(1 1 L)$ scans. The difference patterns in (b) show clear magnetic peaks at half-integer $L$ up to $L=3.5$. No such peak is observed in (c,d) along the $(\frac{1}{2}\frac{1}{2}L)$ and $(11L)$ directions. These observations are consistent with an A-type AFM state, {\it i.e}, the $H = 0$ ground state is such that the intraplane moments are ferromagnetically aligned in the $ab$ plane while the moments in adjacent Eu planes along the $c$~axis are aligned antiferromagnetically. Note that structure-factor calculations for this model indicate $(1 1 L)$ at half-integer values of $L$; we argue that their absence in (d) is due to the form factor of ${\rm Eu}^{2+}$ at these relatively large momentum transfers. (e)~Integrated intensity as a function of temperature $T$ of the (0 0 $\frac{1}{2}$) magnetic Bragg reflection fitted with a power-law to yield $T_{\rm N}=(27.3 \pm 0.8)$~K and $\beta=0.21\pm0.01$. (f) Chemical and A-type AFM ground-state structure of \eag. Neutron-diffraction data are insufficient to determine the in-plane moment directions. Therefore, we arbitrarily show the in-plane moments pointing along the next-nearest-neighbor direction.}
\label{Fig:ND_all}
\end{figure*}

\section{\label{Sec:Results} Results and Discussion}

\subsection{\label{Sec:Neutron} Zero-field neutron diffraction}

Figure~\ref{Fig:ND_all}(a) shows zero-field neutron-diffraction scans along the $(00L)$ direction in reciprocal-lattice units (r.l.u.) at 6~K and 30~K, where reflections at half-integer $L$ values are apparent at \mbox{$T = 6$~K\@.}  For more clarity, Fig.~\ref{Fig:ND_all}(b) shows the difference between these two scans, where within experimental uncertainty, there is no evidence for other reflections associated with a modulated structure along the $c$~axis. We also note that the intensities of the new peaks become weaker at larger $L$ values, roughly following the falloff expected from the magnetic form factor of Eu$^{2+}$. Similar differences [i.e., $I{\rm (6~K)} - I({\rm 30~K)}]$ for scans along $(\frac{1}{2}\frac{1}{2}L)$ and $(11L)$, shown in Figs.~\ref{Fig:ND_all}(c,d), respectively, do not reveal any magnetic peaks.

Qualitatively, these newly-emerging Bragg reflections indicate a doubling of the unit cell along the $c$~axis.  These qualitative observations unequivocally establish that these reflections are associated with A-type AFM ordering with propagation vector $\vec{\tau} = \left(0,0,\frac{1}{2}\right)$, consisting of layers of moments aligned ferromagnetically in the $ab$~plane, with moments in adjacent planes along the $c$~axis aligned antiferromagnetically.

The proposed A-type AFM structure is shown in Fig.~\ref{Fig:ND_all}(f), where adjacent nearest-neighbor FM layers along the $c$~axis are rotated by 180$^\circ$ with respect to each other. The direction of the FM moment within an Eu layer cannot be determined from neutron diffraction alone. Using published values, we obtain good agreement with lattice parameters; however, the peak intensities differ significantly from the calculated values due to strong absorption effects by Eu, which are not accounted for in our calculations.

Nevertheless, we are able to confirm the A-type magnetic structure and obtain an estimate for the Eu ordered magnetic moment $\mu = g\langle S\rangle~\mu_{\rm B} = (6.5 \pm 1)~\mu_{\rm B}$ at $T = 6$ K by calculating the magnetic and chemical structure factors, where $S$ is the spin magnetic quantum number, $g$ is the spectroscopic-splitting factor, and $\mu_{\rm B}$ is the Bohr magneton. We note that the large uncertainty in the evaluation of the ordered magnetic moment is mainly due to strong-absorption effects which were not accounted for.  Within the error, the fitted value of $\mu$ agrees with the expected value $\mu = 7\,\mu_{\rm B}$/Eu using $g=2$ and $S=7/2$.

Figure~\ref{Fig:ND_all}(e) shows the integrated intensity of the \mbox{(0 0 $\frac{1}{2}$)} magnetic peak as a function of temperature where we use a weighted power-law function by a Gaussian distribution of $T_{\rm N}$
\bea
I_{\rm (0\,0\, 0.5)}(T) = C|1-T/T_{\rm N}|^{2\beta} \propto \mu^2,
\label{Eq:power-law fit}
\eea
yielding $T_{\rm N} = (27.3 \pm 0.8$)~K and \mbox{$\beta = 0.21 \pm 0.01$.} The $T_{\rm N}$ is in good agreement with the value \mbox{$T_{\rm N}=(27.5 \pm 0.5)$~K} obtained from the $\chi(T)$ and $C_{\rm p}(T)$ measurements below.

\subsection{\label{Sec:MagSus} Magnetic Susceptibility}

\begin{figure}
\includegraphics [width=3in]{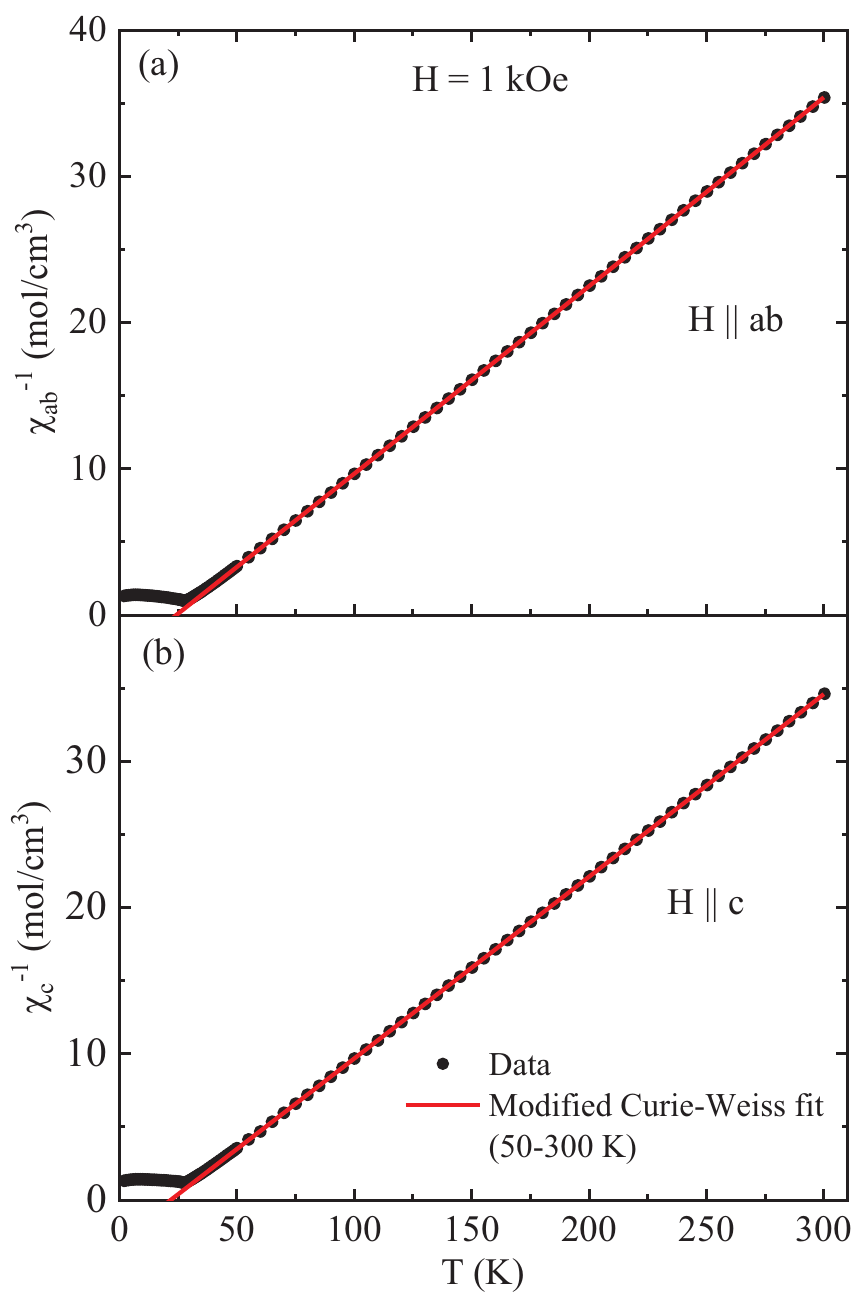}
\caption {Inverse magnetic susceptibility as a function of temperature $\chi^{-1}(T)$ measured for $H = 1$~kOe, when (a) $H \parallel ab$ and (b) $H \parallel c$.}
\label{Fig_inverse_susceptibility}
\end{figure}

The inverse magnetic susceptibility $\chi^{-1}(T)$ data measured under an applied field $H = 1$~kOe for both $H\parallel ab$ and $H\parallel c$ are shown in Figs.~\ref{Fig_inverse_susceptibility}(a) and \ref{Fig_inverse_susceptibility}(b), respectively. The data for $T \geq 50$~K for both field directions were fitted by the modified Curie-Weiss law
\bea
\chi_{\alpha}(T) =\chi_0 + \frac{C_{\alpha}}{T-\theta_{\rm p\alpha}} \quad (\alpha ~=~ab,~c),
\label{Eq.ModCurieWeiss}
\eea
where $\chi_0$ is the temperature-independent contribution, $C_\alpha$ is the Curie constant, and $\theta_{\rm p}$ is the paramagnetic Weiss temperature. The Curie constant $C_\alpha$ is given by
\bse
\label{Eqs:Calpha_mueff}
\be
C_{\alpha}=\frac{N_{\rm A} {g_\alpha}^2S(S+1)\mu^2_{\rm B}}{3k_{\rm B}} = \frac{N_{\rm A}\mu^2_{\rm {eff, \alpha}}}{3k_{\rm B}},
\label{Eq.Cvalue1}
\ee
where $N_{\rm A}$ is Avogadro's number and the effective magnetic moment is given by
\be
\mu_{\rm {eff, \alpha}} = g_\alpha \sqrt{S(S+1)}\,\mu_{\rm B}.
\label{Eq.mueff}
\ee
\ese
The fits of the $\chi_\alpha^{-1}(T)$ data by Eq.~(\ref{Eq.ModCurieWeiss}) is depicted in Figs.~\ref{Fig_inverse_susceptibility}(a) and \ref{Fig_inverse_susceptibility}(b) for $H\parallel ab$ and $H\parallel c$, respectively, and the fitted parameters are listed in Table~\ref{Tab.chidata}. The effective moments are close to the value 7.94~$\mu_{\rm B}$/Eu  expected for Eu$^{2+}$ spins with $S = 7/2$ and $g = 2$. The positive values of the Weiss temperatures $\theta\rm_{p\alpha}$ are consistent with the \mbox{A-type} AFM order revealed by the above zero-field neutron-diffraction measurements, where the in-plane FM interactions between the Eu spins are dominant over the interplane AFM interactions.

\begin{table}
\caption{\label{Tab.chidata} The obtained Parameters from the fits of the data in Figs.~\ref{Fig_inverse_susceptibility}(a) and~\ref{Fig_inverse_susceptibility}(b) by Eq.~(\ref{Eq.ModCurieWeiss}).  Listed parameters are the $T$--independent contribution to the magnetic susceptibility $\chi_0$, Curie constant per mol $C_\alpha$ in $\alpha = ab, c$ directions, effective moment per Eu $\mu{\rm_{eff}(\mu_B/Eu)} \approx \sqrt{8C}$ and Weiss temperature $\theta\rm_{p\alpha}$ obtained from the $\chi^{-1}(T)$ versus $T$ data for $H = 1$~kOe.}
\begin{ruledtabular}
\begin{tabular}{ccccc}	
Field  & $\chi_0$ 				& $C_{\alpha}$ 		    &  $\mu_{\rm eff\alpha}$ 	& $\theta_{\rm p\alpha}$ \\
direction	& $\rm{\left(10^{-4}~\frac{cm^3}{mol}\right)}$	 & $\rm{\left(\frac{cm^3 K}{mol}\right)}$    & ($\mu_{\rm B}$/Eu)& (K)  \\
\hline
H $\parallel ab$ 		& $-2.6(3)$		&  	7.86(1)	&	7.93(1)		& 24.26(7)   \\
H $\parallel c$ 	    		& $-1.9(3)$ 		&  	7.99(1)	&	7.99(1)	    	& 21.86(7)	\\
\end{tabular}
\end{ruledtabular}
\end{table}

\begin{figure}
\includegraphics [width=3in]{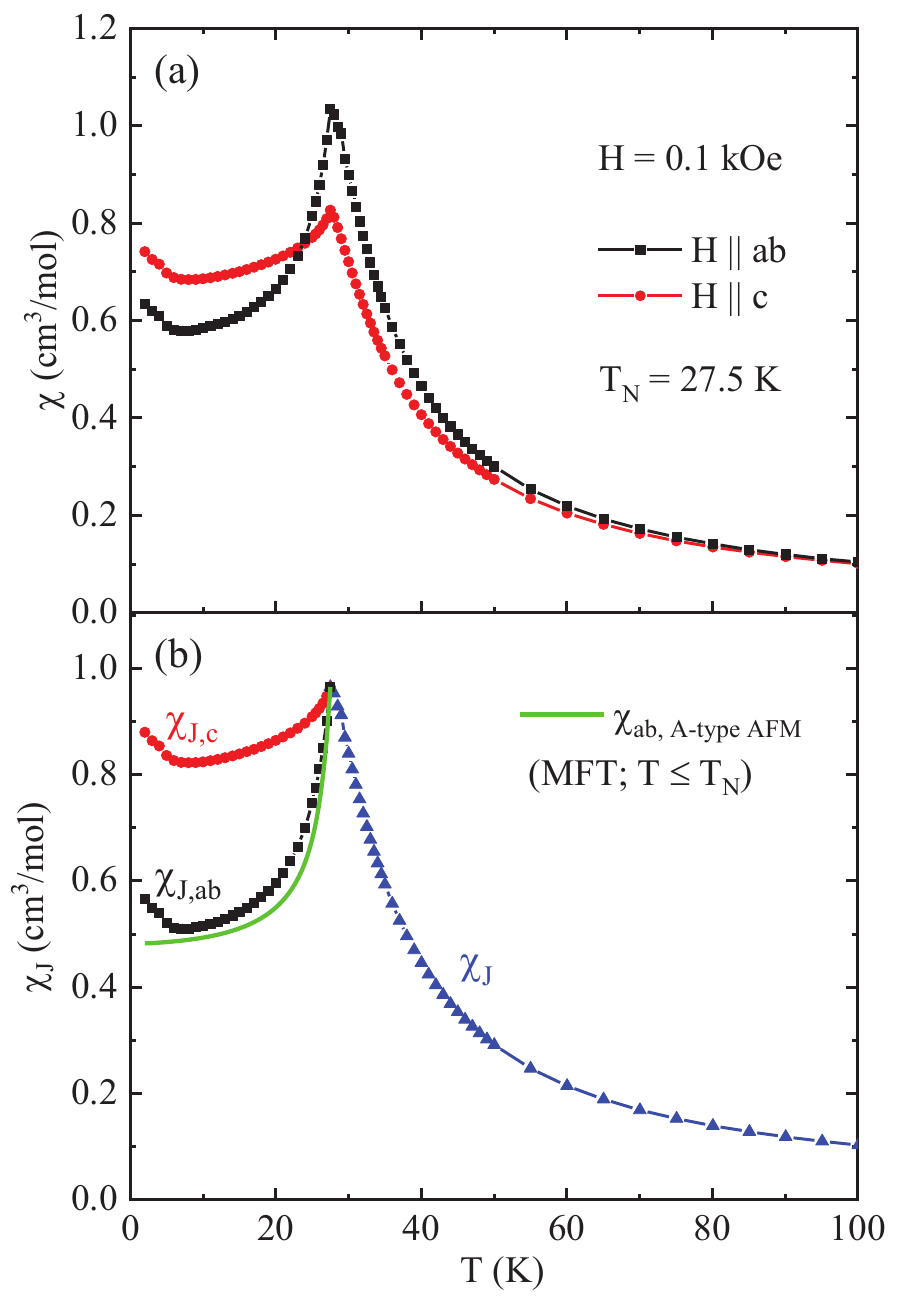}
\caption {(a) Temperature dependence of the magnetic susceptibilities measured for $H = 0.1$~kOe with $H\parallel ab$ (black squares) and $H\parallel c$ (red circles).  The upturns in the $\chi_{ab}(T)$ and $\chi_{c}(T)$ data below  $\sim 5$~K may be associated with an additional  magnetic ordering of unknown type. (b)~Spherically-averaged Heisenberg magnetic susceptibility $\chi_J(T)$ in the PM state with \mbox{$T \geq T_{\rm N}$} obtained using Eq.~(\ref{eq:Chi_ave_PM}) (filled blue triangles). The blue curve connects the data points. The $\chi_{ab}(T)$ and $\chi_{c}(T)$ data in~(a) for $T\leq T_{\rm N}$ are respectively shifted vertically to match the values at $T_{\rm N}$ to the value $\chi_J(T = T_{\rm N})$ = 0.96~cm$^3$/mol.  The $\chi_{J,ab}(T\leq T_{\rm N})$ for A-type AFM order predicted by Eqs.~(\ref{Eqs:Chixy}) for $kd=\pi$~rad and $f = \theta_{\rm p\,ave}/T_{\rm N} = 0.853$ is shown as the green curve.  For A-type ordering with the moments aligned in the $ab$~plane, one theoretically expects $\chi_{J,ab}(0~{\rm K})/\chi_J(T_{\rm N})=1/2$, close to the observed value. }
\label{Fig_Chi-T_0p1kOe}
\end{figure}

The $T$~dependences of the magnetic susceptibilities $\chi$ of \eag\ measured in $H = 0.1$~kOe for the in-plane ($H\parallel ab$) and out-of-plane ($H\parallel c)$ field directions are shown in Fig.~\ref{Fig_Chi-T_0p1kOe}(a). A sharp AFM transition is observed at $T_{\rm N} = 27.5(5)$~K, which is the same as reported earlier for  polycrystalline \eag~\cite{Kranenberg2000}. The anisotropy between $\chi_{ab}$ and $\chi_c$ above $T_{\rm N}$ likely arises from a combination of magnetic-dipole and magnetocrystalline interactions.  The $\chi_J(T)$ data above $T_{\rm N}$ for Heisenberg interactions in the absence of anisotropy are obtained as the average
\bea
\chi_J(T\geq T_{\rm N}) = \frac{1}{3}[2\chi_{ab}(T)+\chi_c(T)]
\label{eq:Chi_ave_PM}
\eea
which is plotted in Fig.~\ref{Fig_Chi-T_0p1kOe}(b).  Then the data at $T\leq T_{\rm N}$ are shifted vertically until they match the $\chi_J(T\geq T_{\rm N})$ data at $T_{\rm N}$ as shown.

The $\chi_{J,ab}$ data strongly decrease on cooling from $T_{\rm N}$ to $T \sim 5$~K, whereas the out-of-plane susceptibility $\chi_{J,ab}$ is less dependent on the temperature, signifying that the $ab$~plane is the easy plane. This observation is in good agreement with the neutron-diffraction results revealing the A-type nature of the magnetic ground state with the moments aligned in the $ab$~plane. However, below $\sim 5$~K, both $\chi_{J,c}$ and $\chi_{J,ab}$ increase sharply,  indicating the occurrence of an additional magnetic transition of unknown nature at $T \sim 5$~K\@. Our neutron-diffraction measurements could not examine the additional transition as their low-$T$ limit was 6~K\@.

Here we utilize the molecular field theory (MFT)~\cite{Johnston2012, Johnston2015} for $c$-axis helical antiferromagnets with the moments aligned in the $ab$~plane with $c$-axis propagation vector~$k$ and interlayer spacing~$d$ for which $kd$ is the turn angle between moments in adjacent layers.  The in-plane magnetic susceptibility $\chi_{Jab}(T)$ associated with Heisenberg spins and spin interactions~$J$ for $T\leq T_{\rm N}$ and no anisotropy can be written as
\bse
\label{Eqs:Chixy}
\be
\frac{\chi_{J,ab}(T \leq T_{\rm N})}{\chi_J(T_{\rm N})}=  \frac{(1+\tau^*+2f+4B^*)(1-f)/2}{(\tau^*+B^*)(1+B^*)-(f+B^*)^2},
\label{eq:Chi_plane}
\ee
where
\be
f=\theta_{\rm p\,ave}/T_{\rm N},
\ee
\be
B^*=  2(1-f)\cos(kd)\,[1+\cos(kd)] - f,
\label{eq:Bstar}
\ee
\be
t =\frac{T}{T_{\rm N}},\quad \tau^*(t) = \frac{(S+1)t}{3B'_S(y_0)}, \quad y_0 = \frac{3\bar{\mu}_0}{(S+1)t},
\ee
the ordered moment versus $T$ in $H=0$ is denoted by $\mu_0$, the reduced ordered moment $\bar{\mu}_0 = \mu_0/\mu_{\rm sat}$ with \mbox{$\mu_{\rm sat} = gS\mu_{\rm B}=7\,\mu_{\rm B}$~here} is determined by numerically solving the self-consistency equation
\be
\bar{\mu}_0 = B_S(y_0),
\label{Eq:barmuSoln}
\ee
$B'_S(y_0) = [dB_S(y)/dy]|_{y=y_0}$, and the Brillouin function $B_S(y)$ is
\be
B_S(y)= \frac{1}{2 S}\left\{(2S+1){\rm coth}\left[(2S+1)\frac{y}{2}\right]-{\rm coth}\left(\frac{y}{2}\right)\right\}.
\ee
\ese

Using the value of $f$ calculated from the values of $\theta_{\rm p,ave}$ and $T_{\rm N}$ from Table~\ref{Tab.chidata}, the calculated $\chi_{J,ab}(T)$ for $T \leq T_{\rm N}$ is shown by the green curve in Fig.~\ref{Fig_Chi-T_0p1kOe}(b).  As seen in the figure, the calculated curve deviates somewhat from the experimental $\chi_{J,ab}(T)$ data, likely due to the additional higher-$T$ magnetic precursor contributions of the anticipated low-$T$ order below 5~K\@. According to the MFT~\cite{Johnston2012, Johnston2015}, at $T=0$ we have
\be
\label{Eq:kd}
\frac{\chi_{J,ab}(T=0)}{\chi_{J,ab}(T_{\rm N})}=\frac{1}{2[1+2~{\rm cos}(kd)+2~{\rm cos}^2(kd)]}.
\ee
Thus, for an A-type AFM, where the turn angle between adjacent $ab$-plane FM layers is $kd\to180^\circ$, one expects \mbox{$\chi_{J,ab}(T=0)/\chi_{J,ab}(T_{\rm N}) \to 1/2$}, close to the value in Fig.~\ref{Fig_Chi-T_0p1kOe}(b).

\begin{figure}
\includegraphics [width=3in]{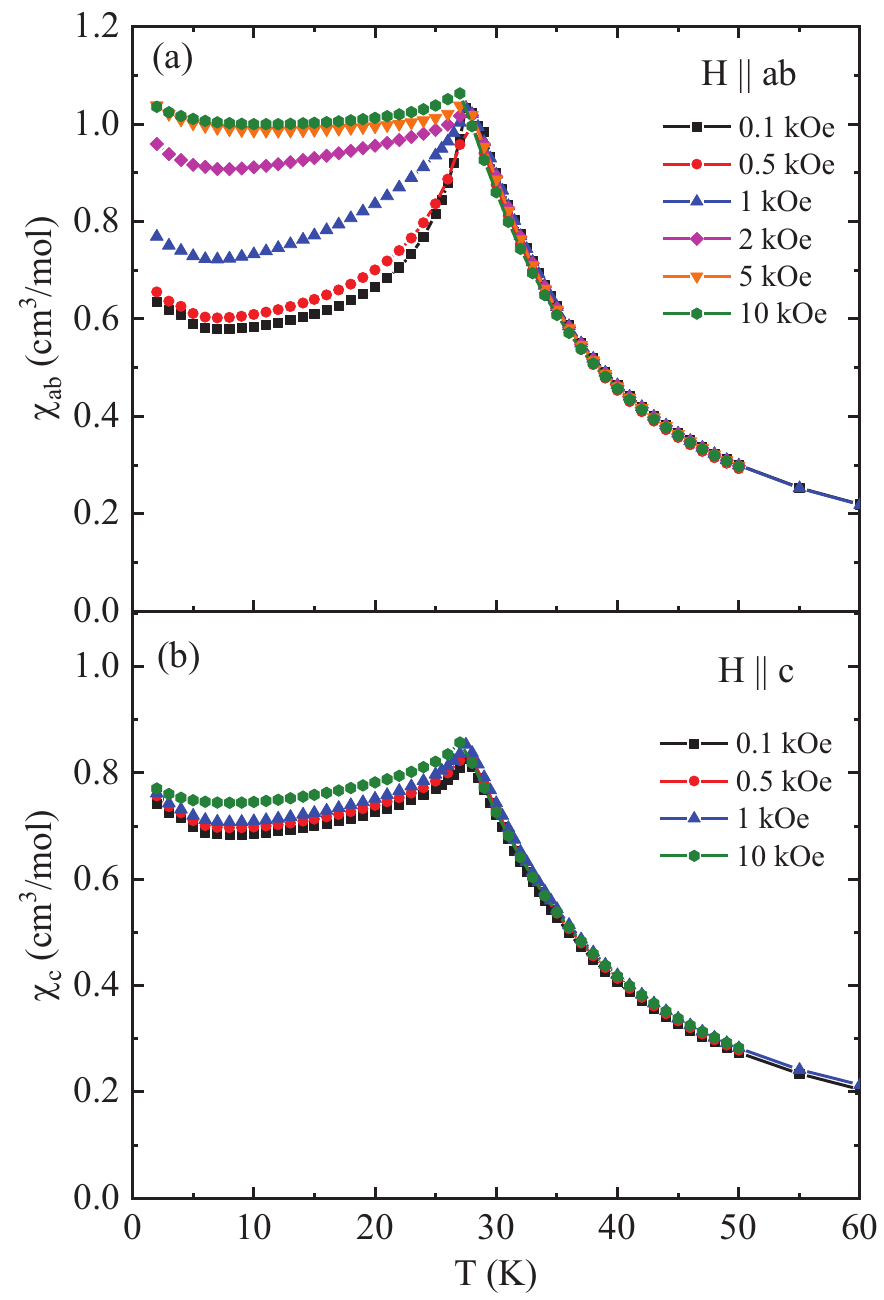}
\caption {Magnetic susceptibility $\chi_\alpha(T)$, $\alpha = ab, c$, at different applied magnetic fields for (a) $H \parallel ab$ and (b) $H \parallel c$. Although, the $\chi_c(T)$ is weakly dependent on $H$ below $T \leq T_{\rm N}$, $\chi_c(T)$ is strongly $H$-dependent up to $H = 10$~kOe.}
\label{Fig_Chi-T_diff_fields}
\end{figure}

The $\chi(T)$ measured at several applied magnetic fields $H$ are shown in Figs.~\ref{Fig_Chi-T_diff_fields}(a) and \ref{Fig_Chi-T_diff_fields}(b) for $H \parallel ab$ and $H \parallel c$, respectively. Interestingly, although the out-of-plane magnetic susceptibility $\chi_c$ remain almost independent of $H$ for $H \leq 10$~kOe, the in-plane susceptibility $\chi_{ab}$ changes significantly with $H$ for $T < T_{\rm N}$ and $H$ up to 5~kOe. Similar behavior was also observed for the trigonal A-type AFM compounds \emb, \ems, \esa, and tetragonal \eg\ with the moments aligned in the $ab$~plane~\cite{Pakhira2022a, Pakhira2020, Pakhira2021a, Pakhira2022b}. We have argued that the A-type ground state spin structure of these materials consist of three-fold (for trigonal) or four-fold (for tetragonal) AFM domains associated with in-plane magnetic anisotropy. As Eu$^{2+}$ moments with $L = 0$ provide negligible single-ion anisotropy, magnetic dipole interaction and other magnetocrystalline anisotropy energy may play a critical role for the formation of AFM domains in these materials. The $H$-dependent change in the $\chi_{ab}(T)$ behavior is due to the reorientation of the spins with in-plane field $H_{\rm ab}$ up to a critical field $H_{c1}$, where all the spins in different domains become perpendicular to the in-plane applied field direction. The spins tend to align along the field direction for $H > H_{c1}$, as expected for a collinear antiferromagnet.

\subsection{\label{Sec:IsoMag} Isothermal magnetization versus applied magnetic field measurements}

\begin{figure*}
\includegraphics [width=6.6in]{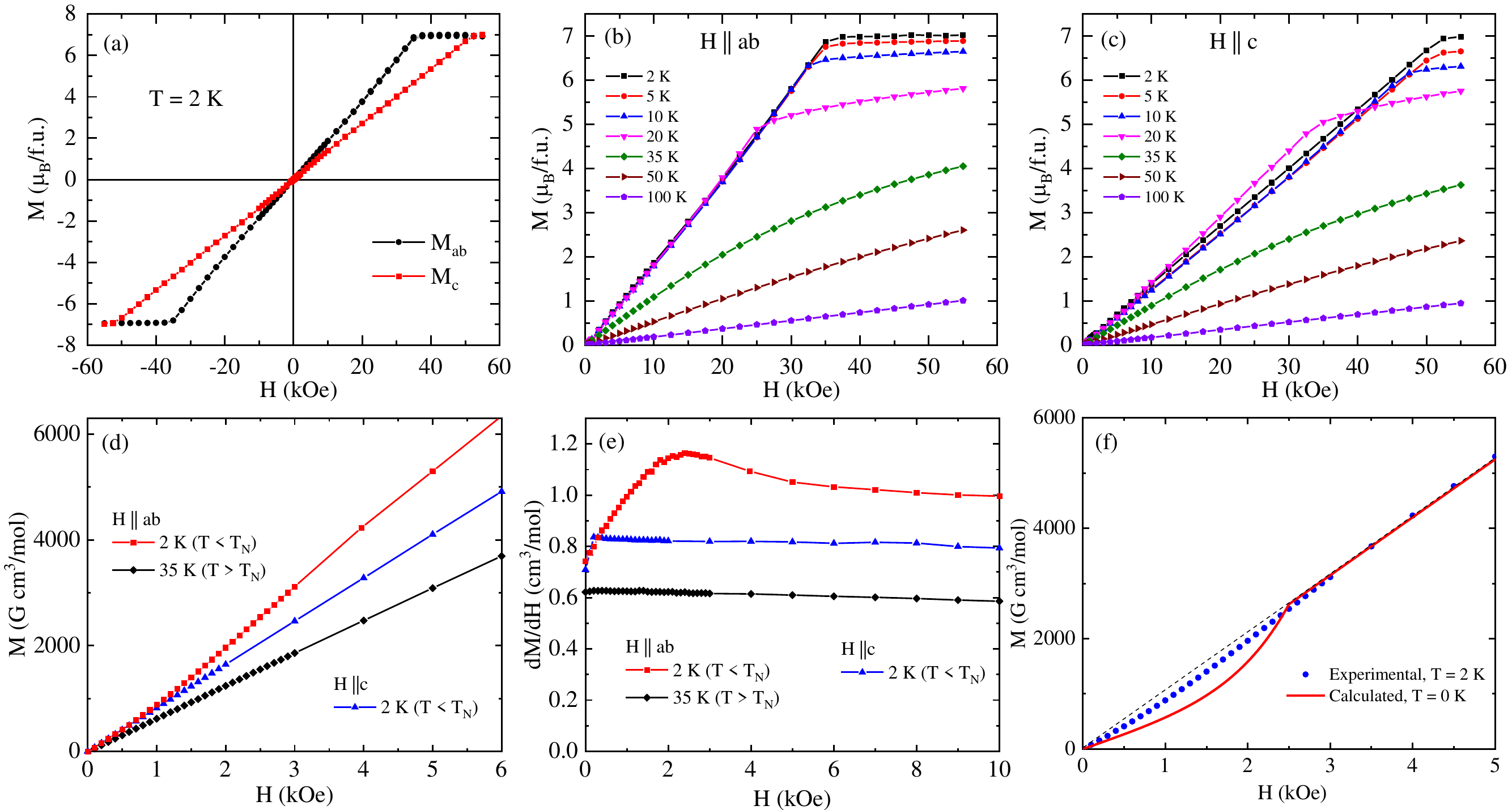}
\caption {(a) Magnetic field dependence of isothermal magnetization $M(H)$ in the hysteresis mode for \mbox{$-5.5~{\rm T} \leq H \leq 5.5$~T} measured at $T = 2$~K for both $H\parallel ab$ and $H\parallel c$. $M(H)$ behavior measured at different temperatures for (b)~$H \parallel ab$ and (c)~$H \parallel c$. (d) Low-field $M(H)$ data showing nonlinearity in the $M_{ab}(H)$ behavior for $T < T_{\rm N}$, whereas $M_{c}(H)$ is linear down to the lowest measured temperature 2~K\@. This nonlinearity is clearly reflected in the $dM/dH$ data shown in (e). (f) The experimental magnetization $M_{ab}(H)$ at $T=2$~K along with the theoretical prediction for $T=0$~K with $H_{\rm c1} \approx 2.5$~kOe. The dashed line is the guide to the eye of the high-field extrapolated linear behavior. The $M_{ab}(H)$ data exhibit positive curvature for $H < H_{\rm c1}$ as predicted by our theory, but the origin of the quantitative difference between experiment and theory is not clear at present.}
\label{Fig_M-H}
\end{figure*}

\subsubsection{Overview}

The evolution of the ground-state spin configuration in \eag\ is further probed by isothermal magnetization versus applied magnetic field  $M(H)$ measurements. The $M(H)$ behavior in the hysteresis mode for \mbox{$-5.5~{\rm T} \leq H \leq 5.5$~T} measured at $T = 2$~K is shown in Fig.~\ref{Fig_M-H}(a). No magnetic hysteresis is observed for fields applied either in the $ab$~plane or along the $c$~axis.   Figures~\ref{Fig_M-H}(b) and~\ref{Fig_M-H}(c) show the $M(H)$ behavior measured at different temperatures for $H \parallel ab$ ($M_{ab}$) and $H \parallel c$ ($M_{c}$), respectively, for our full field range 0--55~kOe.  Here both $M_{ab}$ and $M_{c}$ appear to increase linearly with $H$ and saturate above the respective critical field $H^{\rm c}_{ab} = 37.5(5)$~kOe and $H^{\rm c}_c = 52.5(5)$~kOe with a saturation moment $\mu_{\rm sat} =  7.0(5)\,\mu_{\rm B}$/Eu at $T = 2$~K\@. The measured $\mu_{\rm sat}$ value agrees with $\mu_{\rm sat} = gS\mu_{\rm B} = 7~\mu_{\rm B}$/Eu expected for Eu$^{+2}$ ions with spectroscopic-splitting factor $g = 2$ and spin $S = 7/2$.

The significant difference between the critical-field values for the two field directions indicates the presence of a considerable magnetic anisotropy in the system with $ab$-plane ordering preferred over $c$-axis ordering in the A-type AFM structure, as also observed in the magnetic susceptibility behavior in Fig.~\ref{Fig_Chi-T_diff_fields}.  Figures~\ref{Fig_M-H}(b) and \ref{Fig_M-H}(c) show that the $H^{\rm c}$ values decrease with increase in the temperature for $T < T_{\rm N}$ as expected. The $M(H)$ data measured at $T = 50$~K, greater than $T_{\rm N} = 27.5$~K, are also nonlinear for both the field directions, suggesting the presence of short-range dynamic magnetic correlations in \eag\ above $T_{\rm N}$\@.

\subsubsection{Low-field $M_{ab}(H)$ data}

The $M_{ab}(H)$ data at $T=2~{\rm K} \ll T_{\rm N} = 27.5$~K in Fig.~\ref{Fig_M-H}(a) measured over our maximum field range below $T_{\rm N}$ appear to increase linearly up to $H^{\rm c}_{ab} = 37.5(5)$~kOe above which they saturate.  However, a careful study at low fields revealed that $M_{ab}(H)$ at $T=2$~K  exhibits positive curvature below $H \lesssim 2.5$~kOe as shown in Fig.~\ref{Fig_M-H}(d). The positive curvature is more clearly reflected in the $dM_{ab}/dH$ versus $H$ at $T=2$~K plotted in Fig.~\ref{Fig_M-H}(e) that exhibits a broad peak at $H_{\rm c1} = 2.5(1)$~kOe.   On the other hand, no nonlinearity is observed in the $M_{ab}(H)$ data at $T > T_{\rm N}$ or in the $M_{c}(H)$ data at any temperature. A  similar behavior of $M_{ab}(H)$ was observed by us at $T\approx 2$~K, far below the respective $T_{\rm N}$ for other Eu-based trigonal compounds \emb\ and \ems\ containing triangular Eu layers, as well as for the tetragonal compound \eg\ containing square-lattice Eu layers~\cite{Pakhira2022a, Pakhira2021, Pakhira2021a, Pakhira2022, Pakhira2022b}, where each compound exhibits A-type AFM order with the moments aligned in the $ab$~plane as in \eag.

\subsubsection{Theoretical modeling of the low-field $M_{ab}(H)$ data}

\subsubsection*{a. Overview}

 \begin{figure}
\includegraphics [width=2.25in]{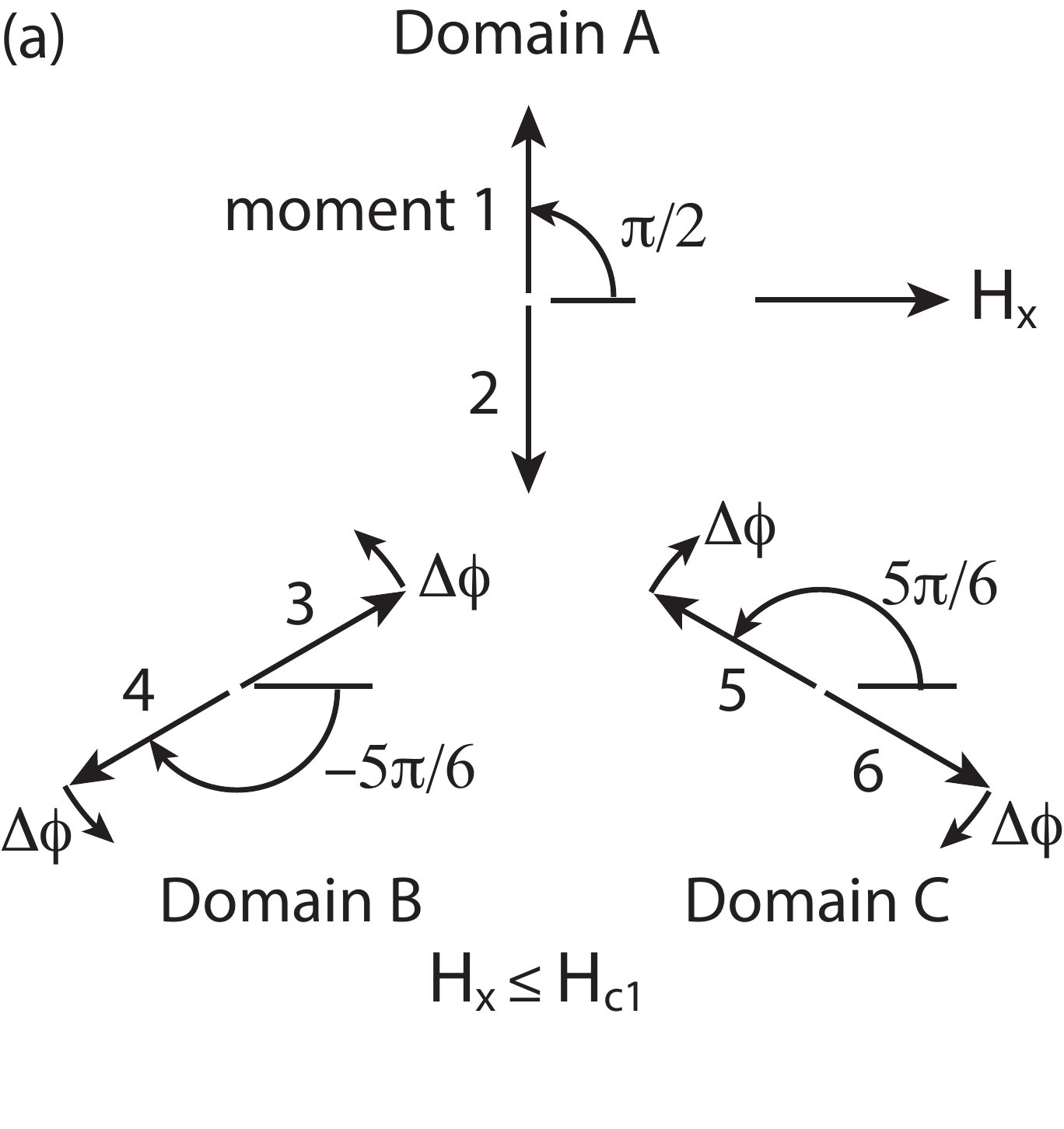}
\includegraphics [width=2in]{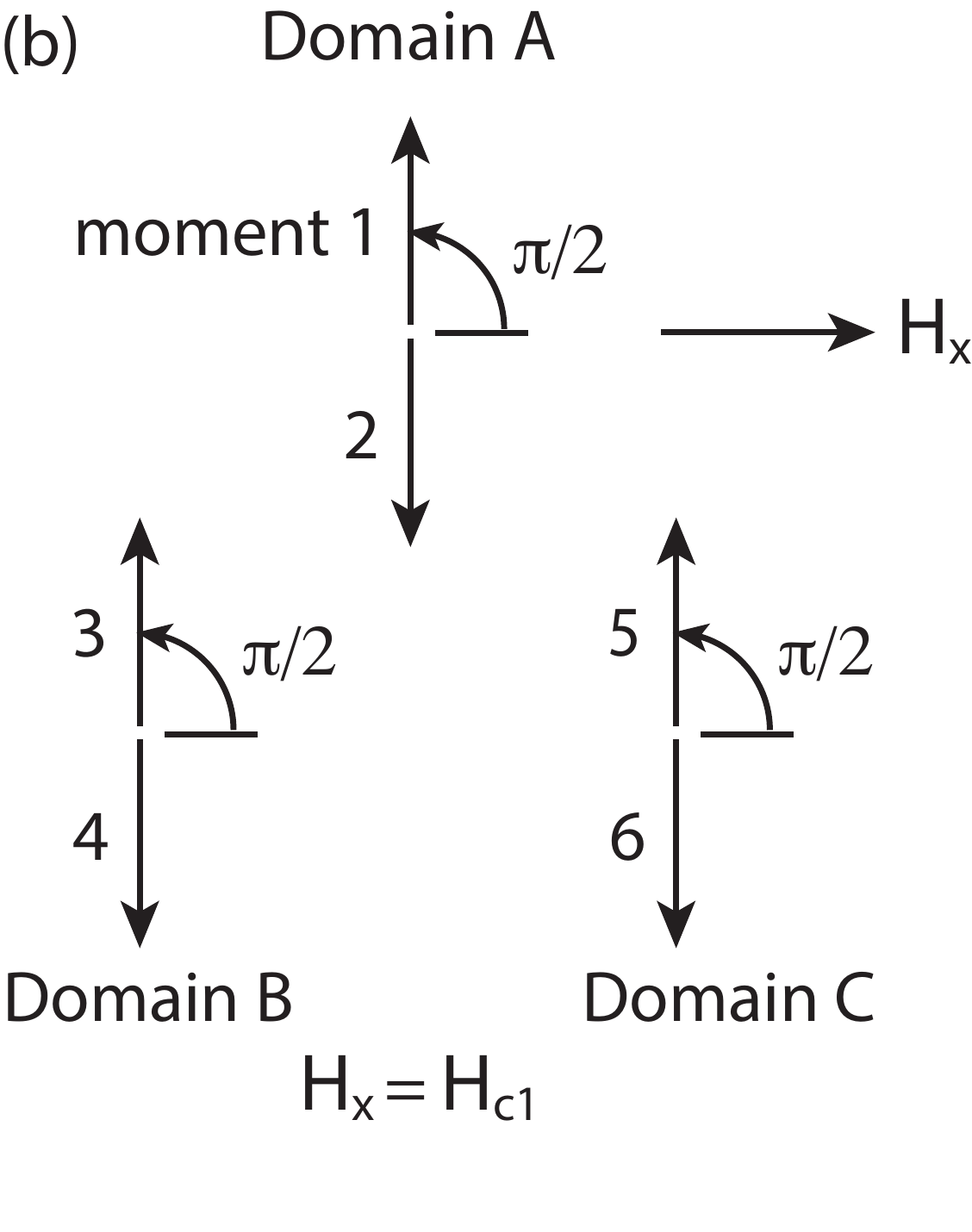}
\caption {(a)~Reorientation of the Eu magnetic moments in the three trigonal $ab$-plane antiferromagnetic domains in a small $ab$-plane magnetic field $H_x<H_{\rm c1}$. Here, the two oppositely-directed arrows in each domain represent the moment orientations in adjacent layers of the A-type AFM structure in small fields.  The arrows indicate the direction and increment $\Delta\phi$ of rotation of the moments in domains B and~C towards the vertical orientation, perpendicular to the applied field ${\bf H}_x$.  The moments in each domain remain antiparallel to each other for $H_x<H_{\rm c1}$ apart from a small canting ($\lesssim 1^\circ$) towards the magnetic field direction that gives rise to the measured magnetization in this field range. (b)~Orientation of the moments at the critical field $H_x=H_{\rm c1}$ where all moments are perpendicular to ${\bf H}_x$ except for the small canting towards ${\bf H}_x$.  At higher fields, all moments cant toward ${\bf H}_x$ for \mbox{$H_{\rm c1} < H_x < H^{\rm c}_{ab}$} until at the critical field $H^{\rm c}_{ab}$ all moments are aligned ferromagnetically in the direction of  ${\bf H}_x$.}
\label{Fig:TrigDomains}
\end{figure}

In order to model the nonlinear low-field $M_{ab}(H)$ data at $T\ll T_{\rm N}$ for \emb, \ems, and \eg, we previously proposed that the A-type AFM ground state of each contains threefold or fourfold A-type AFM domains of moments for the trigonal and tetragonal spin systems, respectively. In the trigonal case, the three domains are associated with a weak $ab$-plane magnetic anisotropy energy
\bea
E_{\rm anis}=K_3\sin(3\phi)
\label{Eq:Eanis}
\eea
with minima in the $ab$-plane azimuthal angle $\phi$ at $\pi/2$, $5\pi/6$ and $-5\pi/6$~rad, where $K_3$ is the anisotropy constant.  Thus in $H=0$, the collinear moments in adjacent layers in \eag\ form three domains with the collinear moments oriented along these three minima as shown in Fig.~\ref{Fig:TrigDomains}(a).

Upon application of $ab$-plane magnetic field ${\bf H}_x$, the antiparallel spins in domains~B and~C initially rotate in a direction to become perpendicular to {\bf H} at $H_{\rm c1}$ as shown by the arrows in Fig.~\ref{Fig:TrigDomains}(a) attached to an angular deviation $\Delta\phi$ for a particular value of the field $H_x$.   This happens because for a collinear antiferromagnet at $T=0$~K, the magnetic susceptibility parallel to the moments is zero, whereas the susceptibility if the moments are perpendicular to the field the magnetic susceptibility  $\chi_\perp=\chi(T){\rm N}$ is nonzero according to molecular-field theory (MFT)~\cite{Johnston2015} and hence the lowest energy occurs if the moments are perpendicular to ${\bf H}_x$, as discussed further below.  With a sufficiently large $H_x\equiv H_{\rm c1}$, all moments are oriented perpendicular to ${\bf H}_x$ apart from a small canting $\lesssim 1^\circ$ towards ${\bf H}_x$ that is responsible for the measured magnetization at this field.  As discussed quantitatively below, the  positive curvature in $M_{ab}(H)$ for $H_x<H_{\rm c1}$ as seen in Fig.~\ref{Fig_M-H}(f) arises from this magnetic-field-induced reorientation of the moments in Domains~B and~C\@.  At fields larger than $H_{\rm c1}$, according to MFT~\cite{Johnston2015} $M_{ab}(H)$ increases linearly up to the critical field $H^{\rm c}_{ab}$ at which all moments are aligned parallel to ${\bf H}_x$ and hence the magnetization saturates to the value $7\mu_{\rm B}$/Eu, in agreement with the experimental data in Fig.~\ref{Fig_M-H}(a).

\subsubsection*{b. Calculations}

Here we summarize the development of the model in Ref.~\cite{Pakhira2022a} for \emb\ and \ems\ as applied to \eag.  In the small fields $0 \leq H_x\leq H_{\rm c1}$, the angles of the moments in domains A, B, and C in Fig.~\ref{Fig:TrigDomains}(a) with respect to the positive $x$~axis are respectively given by
\bea
\phi_{\rm A} &=& \frac{\pi}{2},\nonumber\\
\phi_{\rm B} &=& -\frac{5\pi}{6} + \Delta\phi \quad (0\leq \Delta\phi \leq \pi/3), \label{Eqs:phiABC}\\
\phi_{\rm C} &=& -\frac{\pi}{6} - \Delta\phi.  \quad (0\leq \Delta\phi \leq \pi/3). \nonumber
\eea
The anisotropy energy averaged over the moments in the three domains in the field range $0\leq H_x\leq H_{\rm c1}$ using Eqs.~(\ref{Eq:Eanis}) and (\ref{Eqs:phiABC}) is
\bea
E_{\rm anis\,ave} &=&  - \frac{K_3}{3}[1+2\cos(3\Delta\phi)].\label{Eq:EanisAve}
\eea
The magnetic energy in the regime $0\leq H_x \leq H_{\rm c1}$ is given by
\bse
\bea
E_{\rm mag} &=& -\chi_\perp H_x^2\sin(\phi), \label{Eq:Emag}
\eea
where $\chi_\perp$ is the $ab$-plane magnetic susceptibility at $T=0$~K when all moments are perpendicular to ${\bf H}_x$, {\it i.e.}, when $\phi=\pi/2$.  Summing over the angles of the moments in the three domains in Eq.~(\ref{Eqs:phiABC}) and dividing by $3$ gives the average magnetic energy as
\bea
E_{\rm mag\ ave} = -\frac{\chi_\perp H_x^2}{3}\left[1+2\sin^2\left(\frac{\pi}{6} +\Delta\phi\right) \right].
\label{Eq:EmagAve}
\eea
\ese
The total average energy $E_{\rm ave} = E_{\rm anis\,ave} + E_{\rm mag\,ave}$ is given by the sum of Eqs.~(\ref{Eq:EanisAve}) and (\ref{Eq:EmagAve}).  Then normalizing $E_{\rm mag\ ave}$ by $K_3$ gives
\bea
\frac{E_{\rm ave}}{K_3} &=& -\frac{1}{3}\bigg\{1+2\cos(3\Delta\phi)] \label{Eq:Eave2}\\
&&+\frac{\chi_\perp}{K_3} H_x^2 \left[1+2\sin^2\left(\frac{\pi}{6} +\Delta\phi\right) \right]\bigg\}.\nonumber
\eea

Minimizing $E_{\rm ave}/K_3$ with respect to the quantity $\chi_\perp H_x^2/K_3$ yields the relationship between $\Delta\phi$ and $H_x$ given by
\bse
\bea
3\csc\left(\frac{\pi+6\Delta\phi}{3}\right)\sin(3\Delta\phi) = \frac{\chi_\perp H_x^2}{K_3},
\label{Eq:DeltaPhiHx}
\eea
which yields
\bea
\frac{\chi_\perp H_x^2}{K_3}(\Delta\phi=0) &=&0,\\
\frac{\chi_\perp H_{{\rm c1}}^2}{K_3}(\Delta\phi=\pi/3) &=& 9/2.\label{Eq:FindK3}
\eea
\ese

Equation~(\ref{Eq:FindK3}) allows the anisotropy constant $K_3$ in \eag\  to be calculated from the known values of the molar $\chi_\perp=\chi_J(T_{\rm N}) = 0.96$~cm$^3$/mol from Fig.~\ref{Fig_Chi-T_0p1kOe}(b) and $H_{\rm c1} = 2.5$~kOe according to
\bea
K_3 = \frac{\chi_\perp H_{\rm c1}^2}{(9/2)N_{\rm A}} = 1.4\times10^{-3}~{\rm meV/Eu},
\label{Eq:K3eag}
\eea
where $N_{\rm A}$ is Avogadro's number.  For comparison, \mbox{$K_3 = 6.5\times10^{-5}$~meV/Eu} in trigonal \emb~\cite{Pakhira2022a}, $K_3 = 1.8\times10^{-5}$~meV/Eu in trigonal \ems~\cite{Pakhira2022a}, and $K_4=1.4\times 10^{-3}$~meV/Eu in tetragonal \eg~\cite{Pakhira2022b}.

For $0\leq H_x\leq H_{\rm c1}$,  the magnetization $M_x$ of the collinear moments in a domain at $T=0$ versus $H_x$ only arises from the component of {\bf M} perpendicular to the ferromagnetically-aligned layers in the A-type AFM structure, because the parallel component gives no contribution at $T=0$~K\@.  The normalized magnetization averaged over the three domains using Eqs.~(\ref{Eqs:phiABC}) is
\bea
\frac{M_{x\,\rm ave}(\Delta\phi)}{M_x(H_{\rm c1})} &=& \frac{1}{3}\left[1+2\sin^2\left(\frac{\pi}{6}+\Delta\phi\right)\right].
\eea
Solving for $\Delta\phi(H_x)$ using Eq.~(\ref{Eq:DeltaPhiHx}) and the known values of $K_3$ and $M_x(H_{\rm c1})$, a plot of $M_{x\,\rm ave}$ versus $H_x$ over the range $0\leq H_x\leq H_{\rm c1}$ is shown in Fig.~\ref{Fig_M-H}(f).  At higher fields $H_{\rm c1}\leq H\leq H^{\rm c}_{ab}$, one has $M(H_x)=\chi_\perp H_x$, above which the magnetization saturates.

\subsection{\label{Sec:Heatcap} Heat capacity}

\begin{figure}
\includegraphics [width=3.3in]{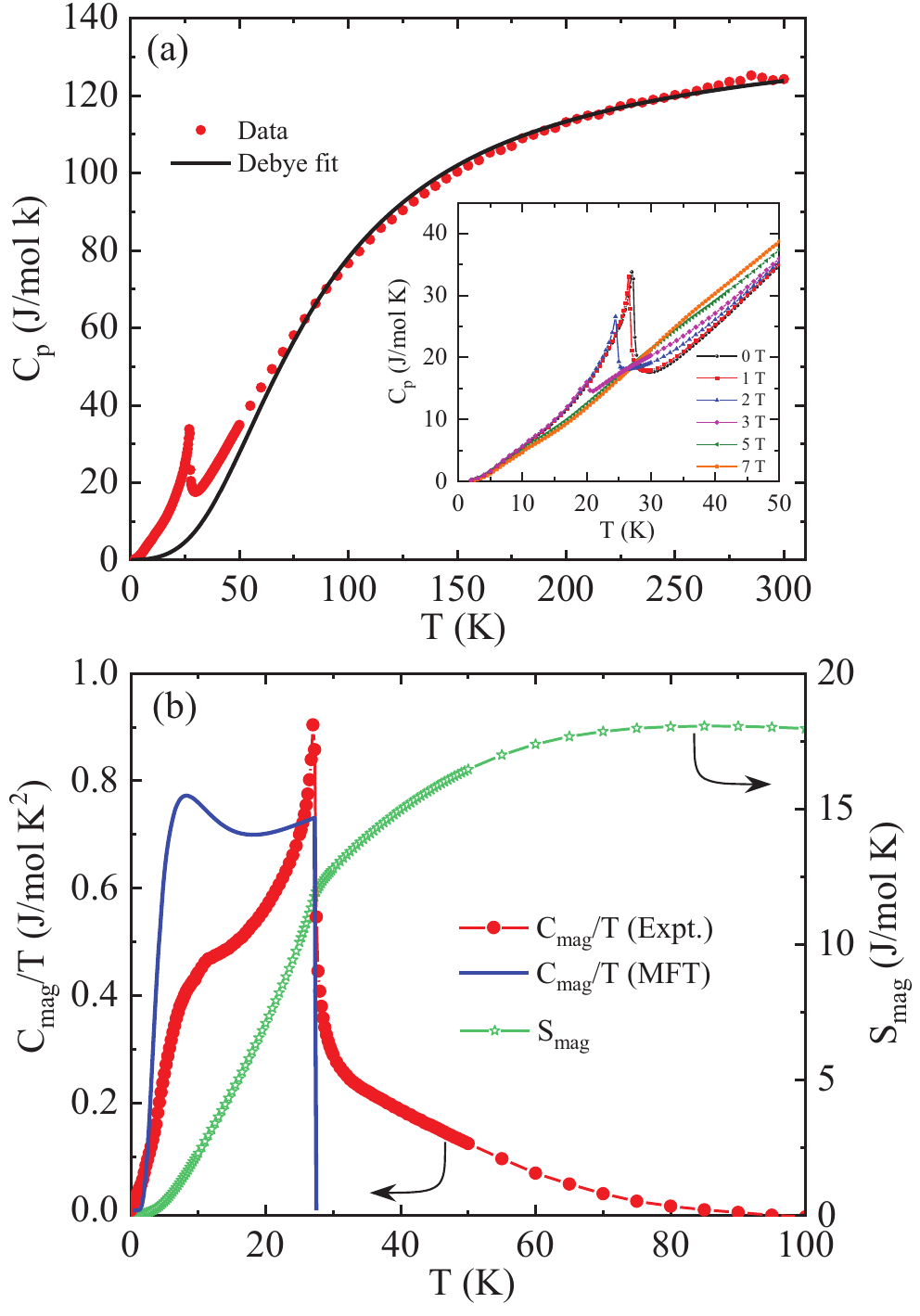}
\caption {(a)~Temperature dependence of the zero-field $C_{\rm p}(T)$ for \eag\ (filled red circles) along with a fit by Eq.~(\ref{Eq:Debye_Fit}) (solid black curve).  Inset: $C_{\rm p}$ vs $T$ in magnetic fields~$H$ from 0 to 7~T\@. (b)~Plot of $C_{\rm mag}/T$ vs ~$T$ in $H=0$ below 100~K  (filled red circles, left ordinate) and the corresponding magnetic entropy $S_{\rm mag}$ vs $T$ (right ordinate) calculated from the $C_{\rm mag}(T)/T$ data using Eq.~(\ref{Smag}). Also shown as a blue curve is $C_{\rm mag}(T)/T$ calculated for $S=7/2$ and $T_{\rm N} = 27.4$~K using the molecular-field theory prediction in  Eq.~(\ref{Eq:deltaCp_T_MFT}). The magnetic entropy $S_{\rm mag}(T)$ calculated using Eq.~(\ref{Smag}) is plotted as the green triangles with the scale on the right ordinate.}
\label{Fig_heat_cap}
\end{figure}

The temperature dependence of the zero-field heat capacity $C_{\rm p}(T)$ of \eag\ is shown in Fig.~\ref{Fig_heat_cap}(a). A clear $\lambda$-type peak is observed in the $C_{\rm p}(T)$ data at \mbox{$T_{\rm N} = 27.5$~K,} indicating the second-order nature of the AFM transition. The peak position shifts to lower temperature with increasing applied field, as shown in the inset of Fig.~\ref{Fig_heat_cap}(a). The $C_{\rm p}(T)$ tends to saturate at a value of $\approx$ 124 J/mol\,K, at $T = 300$~K, close to the classical Dulong-Petit high-$T$ limit $3nR = 124.71$~J/mol\,K, where $n=5$ is the number of atoms per formula unit and $R$ is the molar gas constant.

The molar $C_{\rm p}(T)$ data were fitted by an electronic contribution $\gamma T$ plus the Debye lattice heat-capacity model according to
\bea
C_{\rm p}(T) &=& \gamma T+ n C_{\rm V\,Debye}(T),\label{Eq:Debye_Fit} \\*
C_{\rm V}(T) &=& 9R \left(\frac{T}{\Theta_{\rm D}}\right)^3\int_{0}^{\Theta_{\rm D}/T}\frac{x^4e^x}{(e^x-1)^2} dx,\nonumber
\eea
where $\gamma$ is the Sommerfeld electronic specific-heat coefficient and $\Theta_{\rm D}$ is the Debye temperature. An accurate Pad\'e approximant expression for $C_{\rm V}(T)$~\cite{Goetsch2012} was used for the fit.  The fit is shown by the black curve in Fig.~\ref{Fig_heat_cap}(a), where $\gamma = 21(2)$~mJ/mol\,K$^2$ and \mbox{$\Theta_{\rm D} = 332(2)$~K\@.} The $\gamma$ value is significantly larger than the value of $6(1)$~mJ/mol\,K$^2$ estimated from the theoretical density of states at the Fermi energy $D(E_{\rm F})$ value below.  The enhancement may be due to electron-electron and/or electron-phonon interactions.

Although the AFM ordering temperature of \eag\ is $T_{\rm N} = 27.5$~K, the $C_{\rm p}(T)$ data exhibit a positive deviation from the fit in Fig.~\ref{Fig_heat_cap}(a) for the electronic and lattice contributions up to $~80$~K, indicating the presence of short-range magnetic correlations up to $\sim80$~K\@. The magnetic contribution  $C_{\rm mag}(T)$ to the heat capacity is obtained by subtracting the electronic and lattice contributions from the measured $C_{\rm p}(T)$ data using the above fit and is shown as the red symbols in Fig.~\ref{Fig_heat_cap}(b). The $C_{\rm mag}(T)$ remains finite for $T_{\rm N} < T \lesssim 80$~K due to the presence of short-range dynamic magnetic correlations, in accordance with the observed nonlinear $M(H)$ behavior in Fig.~\ref{Fig_Chi-T_0p1kOe}(a) for $T > T_{\rm N}$ discussed earlier.

In Fig.~\ref{Fig_heat_cap}(b), we have also shown the theoretical $C_{\rm mag}(T)/T$ behavior based on the MFT~\cite{Johnston2015} for this system with $S = 7/2$ as the blue line. According to MFT, the molar $C_{\rm {mag}}(t)$ is given by
\bea
\label{Eq:deltaCp_T_MFT}
C_{\rm {mag}}(t) = R\frac{3S\bar{\mu}_0^2(t)}{(S + 1)t[\frac{(S + 1)t}{3B^{\prime}_S(t)} - 1]},
\eea
where the symbols are defined in Eqs.~(\ref{Eqs:Chixy}). The MFT prediction below~$T_{\rm N}$  in Fig.~\ref{Fig_heat_cap}(b) does not agree well with the data, although the overall shapes below $T_{\rm N}$ are similar. In this regard we must keep in mind the presence of the additional transition below $\sim 5$~K noted above and also the presence of substantial short-range magnetic correlations above $T_{\rm N}$.

The temperature dependence of the magnetic entropy $S_{\rm {mag}}(T)$ is calculated using the experimental data (red symbols) in Fig.~\ref{Fig_heat_cap}(b) and the relation
\bea
\label{Smag}
S_{\rm mag}(T) = \int_{0}^{T}\frac{C_{\rm {mag}}(T)}{T} dT,
\eea
as shown by the green symbols with the scale on the right ordinate of Fig.~\ref{Fig_heat_cap}(b). The $S_{\rm {mag}}(T)$ saturates at $T > 80$~K to a value of $\approx 18$~J/mol~K, which is comparable with the theoretical saturation entropy \mbox{$S_{\rm mag} = R{\rm ln}(2S + 1) = 17.29$ J/mol~K} for Eu$^{2+}$ ions with $S=7/2$. The release of the entropy at  temperatures higher than $T_{\rm N}$ is due to short-range magnetic correlations above $T_{\rm N}$ as indicated from the $C_{\rm mag}(T)/T$ vs~$T$ data in Fig.~\ref{Fig_heat_cap}(b), as also previously found in other Eu- and Gd-based $S = 7/2$ compounds~\cite{Pakhira2020, Pakhira2021a, Pakhira2022, Sangeetha2020, Pakhira2016}.

\subsection{\label{Sec:Resis} Electrical resistivity}

\begin{figure}
\centering
\includegraphics[width=3in]{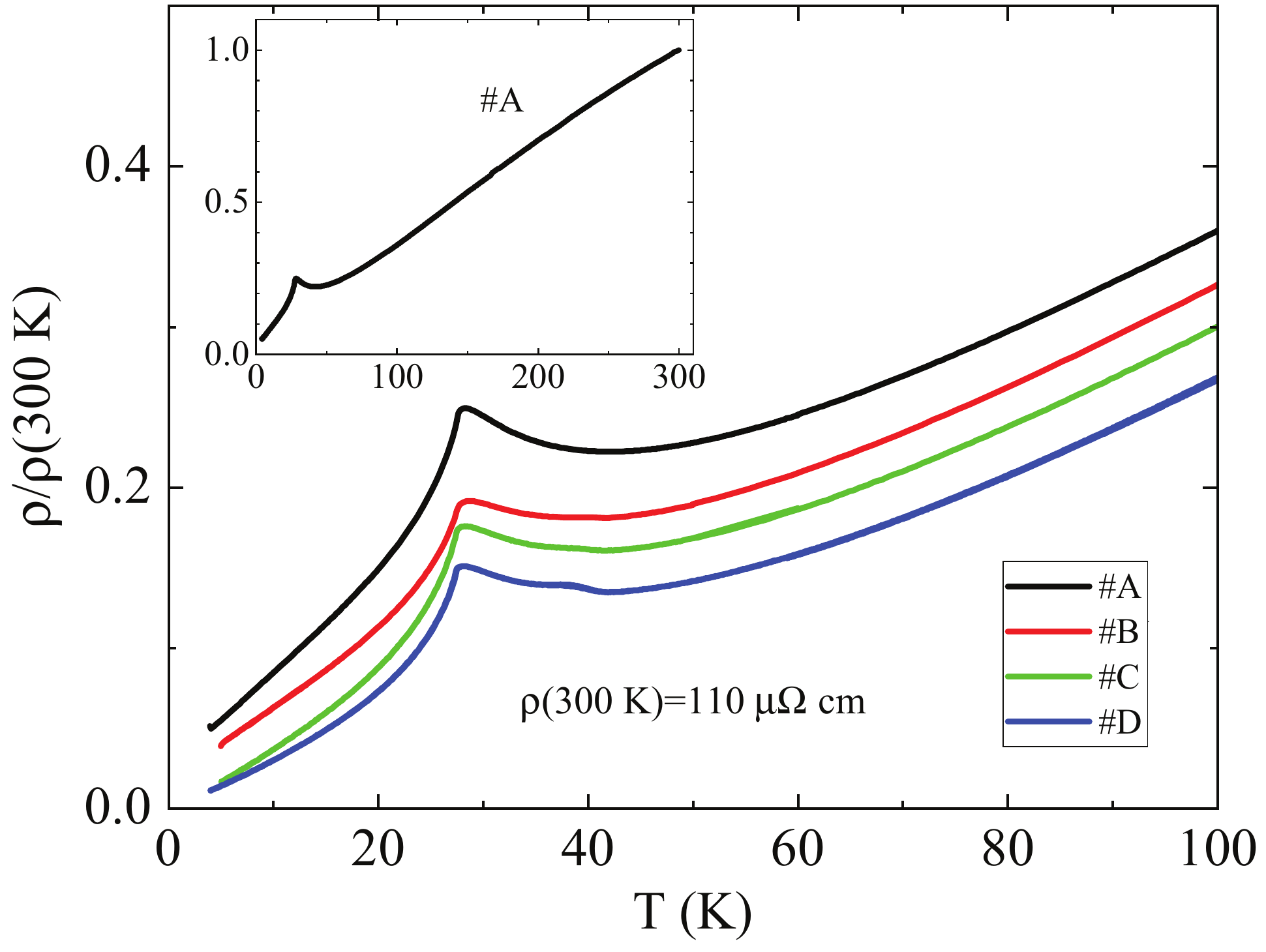}
\caption {Temperature $T$-dependent in-plane electrical resistivity~$\rho$ of four EuAl$_2$Ge$_2$ crystals in $H=0$~T below 100~K\@. The inset shows the full temperature dependence of crystal \#A up to room temperature.}
\label{R(T)H0}
\end{figure}

While the general trend of the electrical resistivity~$\rho$ in the paramagnetic state of EuAl$_2$Ge$_2$ is a metallic decrease on cooling below room temperature as illustrated in the inset of Fig.~\ref{R(T)H0}, anomalous behavior is observed on approaching $T_{\rm N}$ from above as shown in the main panel.  In particular, the resistivity develops significant positive curvature from $\sim 80$~K down to $T_{\rm N}=27$~K, corresponding to the development of dynamic short-range magnetic correlations observed in the heat capacity data in Fig.~\ref{Fig_heat_cap}. Loss of spin-disorder scattering due to long-range AFM ordering  leads to the rapid decrease in the resistivity on cooling below $T_{\rm N}$.

\subsubsection{Electrical resistivity in magnetic fields $H\parallel c$ axis}

\begin{figure}
\centering
\includegraphics[width=3in]{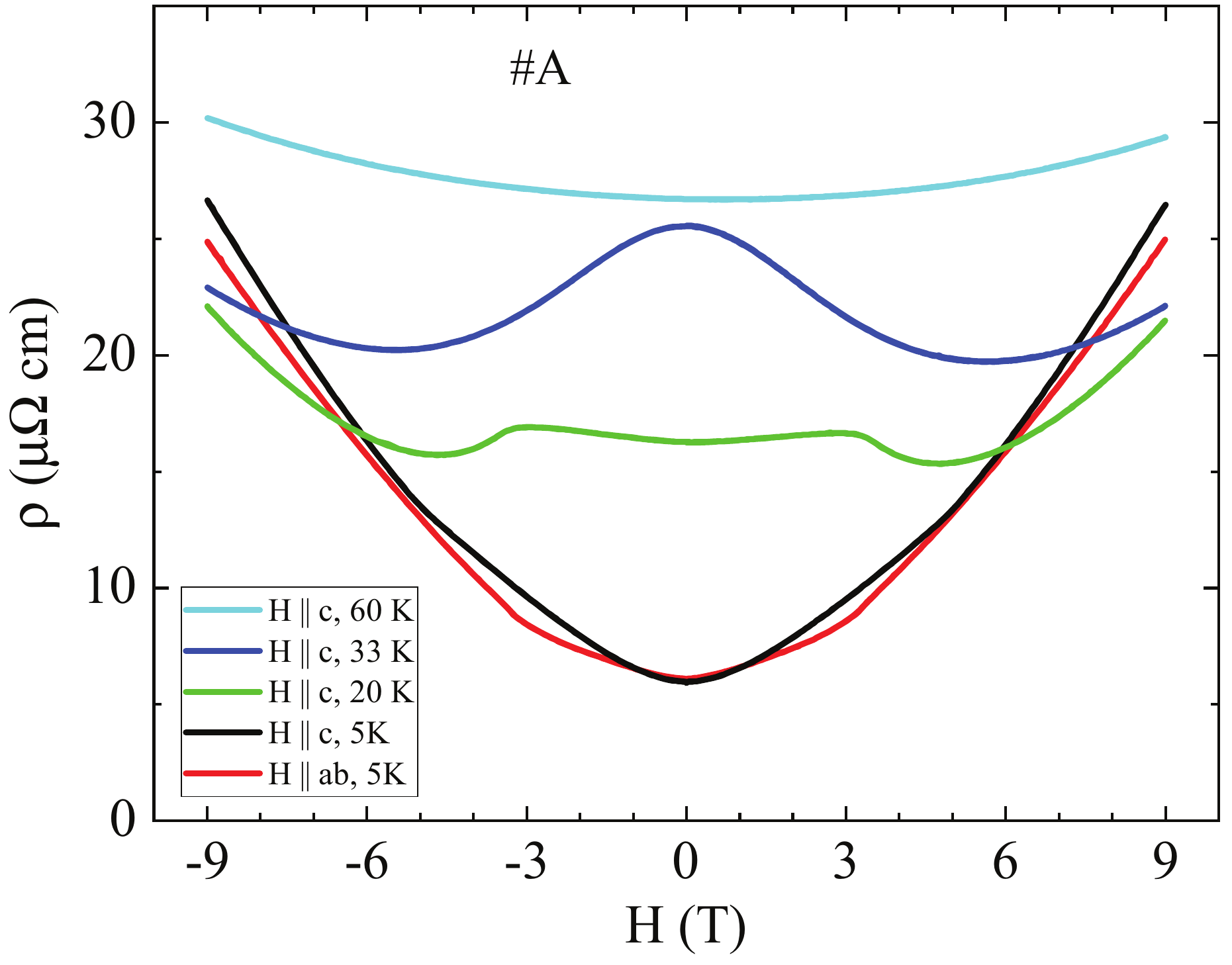}
\caption {In-plane resistivity $\rho$ of EuAl$_2$Ge$_2$ crystal~\#A in magnetic fields in the $H \parallel c$ configuration. Measurements were taken at temperatures of 60~K in the paramagnetic state during the initial development of magnetic correlations (cyan line), at 33~K in the more-correlated paramagnetic state (purple line), and at 20~K (green line) and 5~K (black line) in the A-type AFM state.  For reference we show data taken in the $H \parallel ab$ configuration at 5~K (red line) for which the critical field is about 3.5~T from Fig.~\ref{Fig_M-H}(c).}
\label{resHc}
\end{figure}

In Fig.~\ref{resHc} we show the field-dependent resistivity, measured in magnetic fields parallel to the crystal $c$~axis. Measurements were taken at characteristic temperatures of 60~K (in the paramagnetic state with weak magnetic correlations, cyan line), at 33~K in the correlated paramagnet state (purple line), and at 20~K (green line) and 5~K (black line) in the type-A AFM state.  Magnetization versus field measurements at 5~K and 20~K in this configuration, Fig.~\ref{Fig_M-H}(c) above, show a linear increase at low fields and saturation at fields at about 5~T and 3~T, respectively, in very good agreement with the features seen in the $\rho_a(H_c)$ curves.  At 20~K the resistivity decreases above 3~T, evidencing the suppression of spin-disorder scattering.   At $T=5$~K, the $\rho(H)$ curve shows a slope change at $\sim5$~T\@. For comparison we show resistivity data measured at 5~K in the $H \parallel ab$ configuration, revealing a much clearer feature at the saturation field of $\approx 3.5$~T (red curve in Fig.~\ref{resHc}).  

Note that in the paramagnetic state at 60~K, the resistivity in Fig.~\ref{resHc} increases monotonically with magnetic field, close to the $\rho \sim H^2$ dependence expected for weak-field orbital magnetoresistance~\cite{Ziman2001}. The symmetry of the curve with respect to the sign of the magnetic field suggests minimal contribution of a spurious Hall effect signal in the  resistivity measurements.  In the correlated paramagnet state at 33~K the resistivity decreases with field up to a field of $\sim$ 6~T, due to field-induced suppression of spin-disorder scattering. Positive magnetoresistance is restored in the spin-polarized state above 6~T\@.

\subsubsection{Electrical resistivity in magnetic fields $H\parallel ab$~plane}

\begin{figure}
\centering
\includegraphics[width=3in]{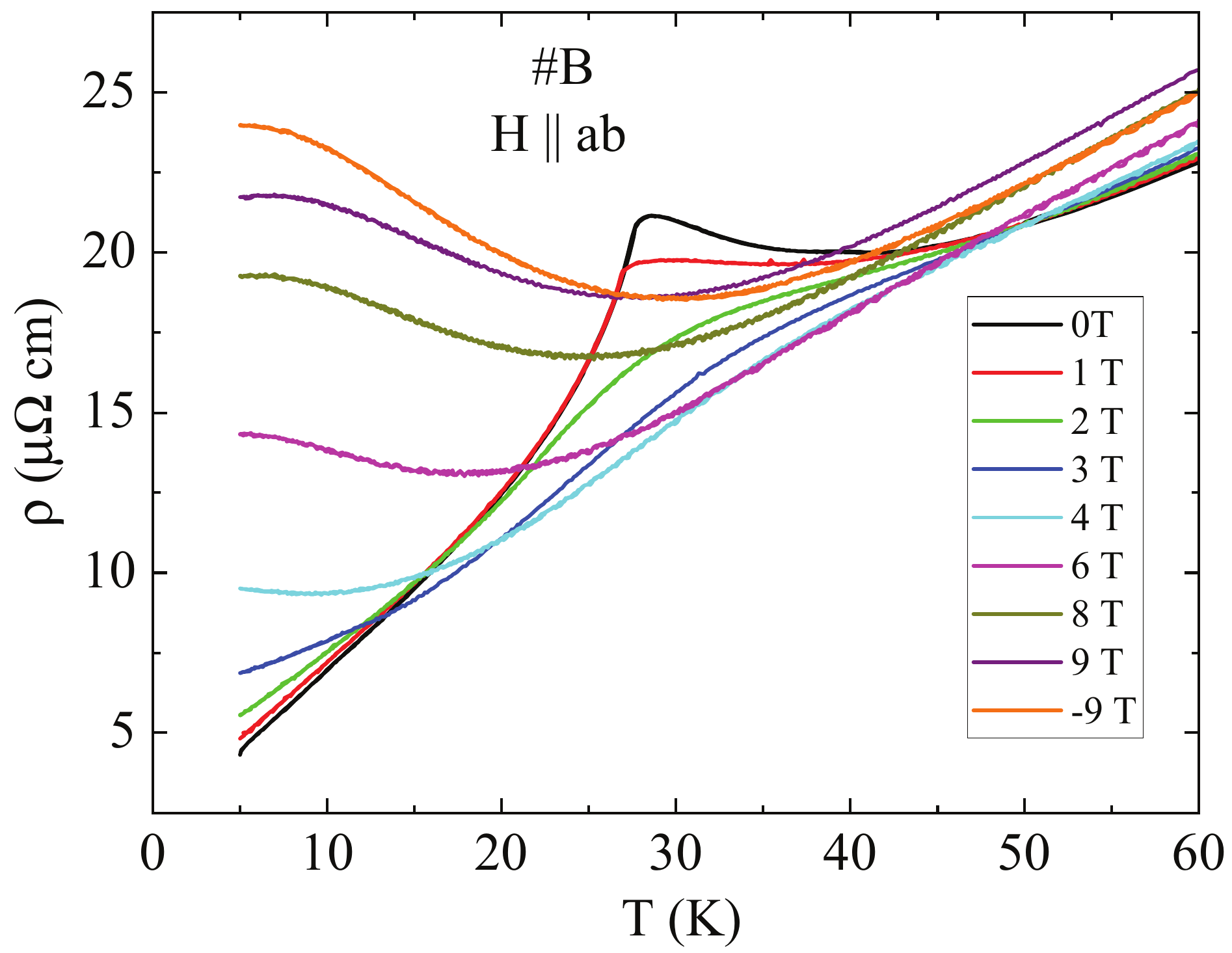}
\caption {Temperature-dependent resistivity of EuAl$_2$Ge$_2$ in magnetic fields $H \parallel ab$. The sharp feature accompanying long-range AFM ordering at $T_{\rm N}=27$~K in zero field moves to somewhat lower temperature in a field of 1~T (red) and smears and moves to higher temperatures in fields of 2~T (green) and 3~T (blue). Measurements in positive and negative fields of 9~T reveal some contamination of the resistivity signal with the Hall voltage, suggesting a sign change of the Hall effect at around 30~K in the 9~T field.}
\label{resHabT}
\end{figure}

In Fig.~\ref{resHabT} we show the evolution of the temperature-dependent resistivity of EuAl$_2$Ge$_2$ with magnetic field applied parallel to the conducting $ab$~plane. This field effectively alters the interplane alignment of the ferromagnetic planes in the type A antiferromagnet with respect to the field, as discussed in Sec.~\ref{Sec:IsoMag}.  A strong enough magnetic field of 1~T (red curve) suppresses the pre-transition resistivity increase and brings the sharp feature observed in zero field at $T_{\rm N}=27$~K to somewhat lower temperatures. With a further field increase to 2~T (green curve), the sharp feature at $T_{\rm N}$ is smeared and transforms into a broad crossover. It shifts to higher temperatures at 3~T (blue curve) and becomes hard to distinguish at higher fields, clearly showing the importance of the spin-polarized state for its observation.  

\begin{figure}
\centering
\includegraphics[width=3in]{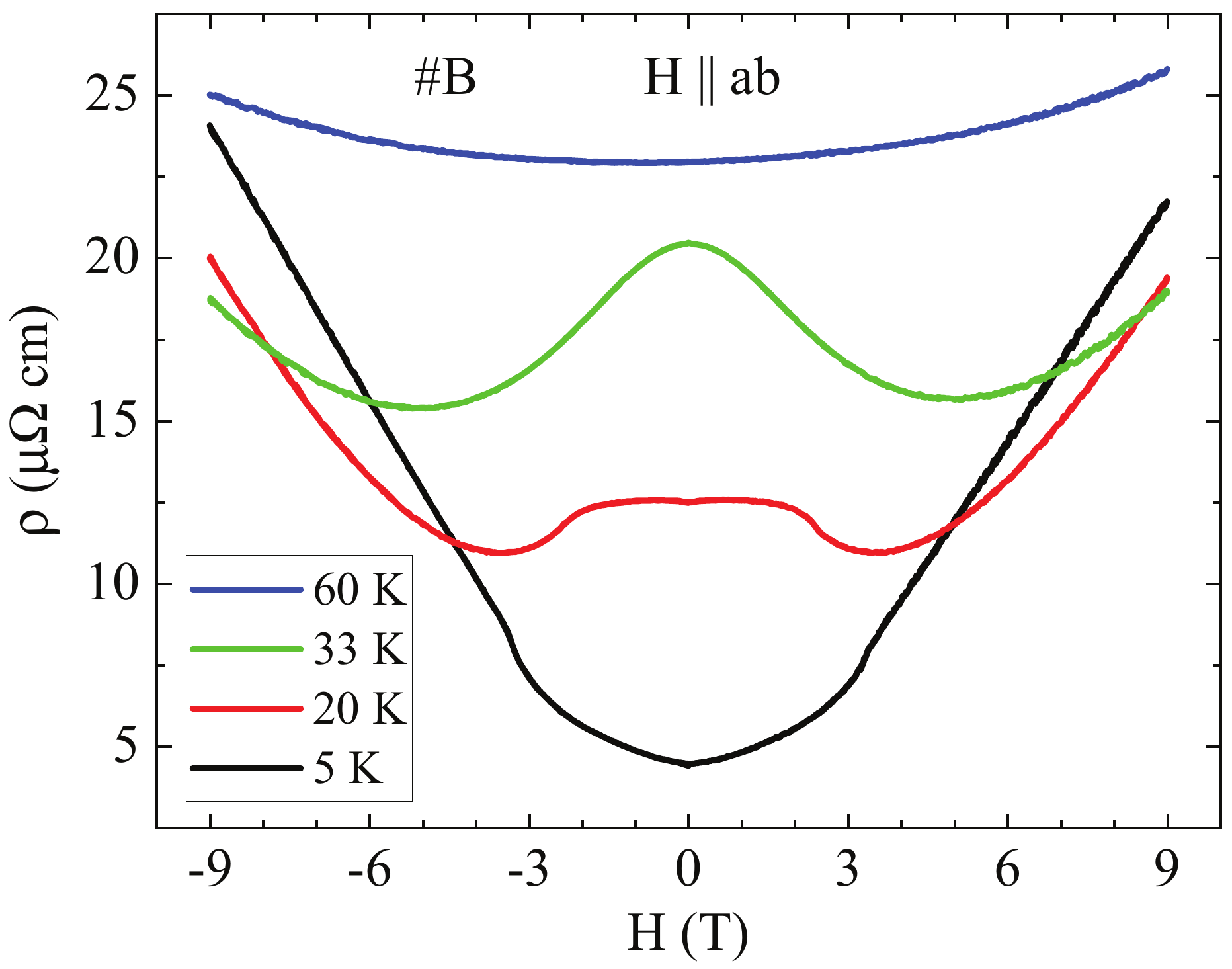}
\caption {In-plane resistivity $\rho$ of EuAl$_2$Ge$_2$ in magnetic fields with the $H \parallel ab$ configuration.  Measurements were taken at characteristic temperatures of 60~K (in the paramagnetic state with weak magnetic correlations, blue line), at 33~K in the correlated paramagnet state (green line), and at 20~K (red line) and 5~K (black line) in the A-type AFM state.}
\label{resHabH}
\end{figure}

Figure~\ref{resHabH} shows the field-dependent resistivity measured in magnetic fields parallel to the sample $ab$~plane. Measurements were taken at characteristic temperatures of 60~K in the paramagnetic state above magnetic correlations development (blue line), at 33~K in the correlated paramagnetic state (green line), and at 20~K (red line) and 5~K (black line) in the A-type AFM state. Magnetization measurements at 5~K and 20~K in this configuration, Fig.~\ref{Fig_M-H}(b), show positive curvature at the lowest fields, zoomed in Fig.~\ref{Fig_M-H}(d), followed by a linear increase and saturation at fields at about 3.5~T and 2.5~T, respectively. This is in very good agreement with the features seen in $\rho(H)$ curves.  At 20~K the resistivity decreases above 2~T, reaches a minimum at 3~T and increases on further field increase.   Note a tiny resistivity increase for the 5~K and 20~K curves, presumably related to magnetic-moment rotations as discussed above in Sec.~\ref{Sec:IsoMag}.

\subsection{\label{Sec:ARPES} Electronic structure from ARPES measurements and DFT calculations}

\begin{figure*}
\centering
\includegraphics[width=6.6in]{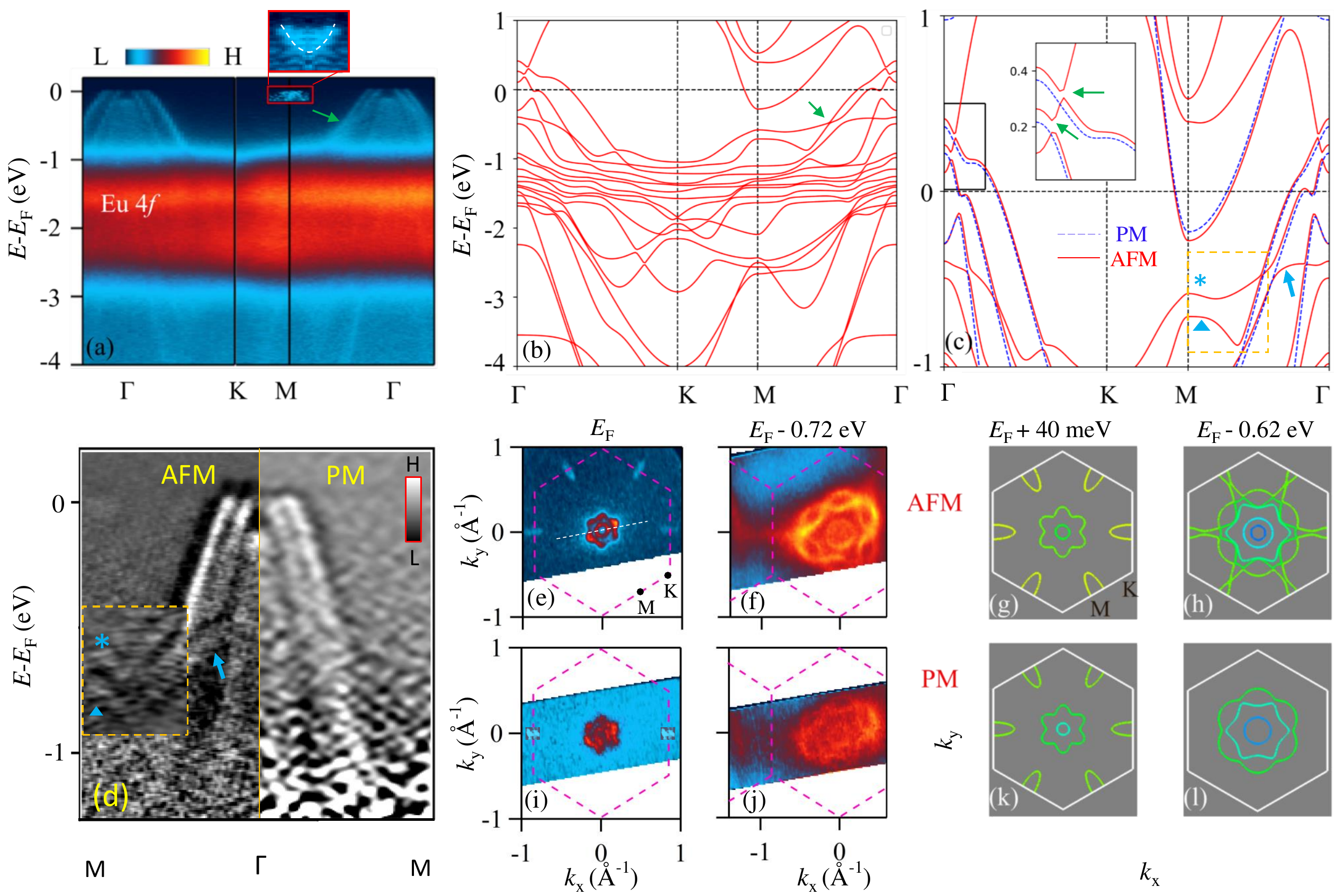}
\caption {Electronic structure of EuAl$_2$Ge$_2$. (a)~ARPES spectrum of EuAl$_2$Ge$_2$ along the ${\rm {\Gamma}-{K}-{M}-{\Gamma}}$ path measured in the AFM phase (9 K) using $h\nu=$ 91 eV ({\it k}$_z$ $\sim$ 0). The inset shows the zoomed-in spectra of the electron pocket at the M point. The arrow indicates the crossing point of two bands. (b)~Theoretical band dispersions including spin-orbit coupling (SOC), Hubbard $U=$ 5 eV, and A-type AFM spin-configuration using DFT. The arrow indicates the crossing of bands. (c)~Theoretical band dispersions in the AFM and PM phases are plotted together. The inset shows zoomed-in spectra around ${\rm \Gamma}$. Band inversion/avoided-crossing features are indicated by the two blue arrows in the inset. Compared to the PM phase, a few extra bands appear in the AFM phase and some of them are indicated by an arrow, star, and triangle symbols. (d)~Two-dimensional second-derivative of the ARPES spectra along ${\rm {\Gamma}-M}$ for AFM and PM phases. Bands within the dashed box are captured by theoretical calculations in~(c). Fermi surface and constant-energy contours for the AFM phase in the experiment [(e)-(f)] and theory [(g)-(h)] and similarly, for the PM phase (40 K) in the experiment [(i)-(j)] and theory [(k)-(l)]. Different energy values are used between the experiment and theory as the position of the Fermi level is slightly different between them.  The ARPES spectra in Fig.~\ref{TD} were taken along the cut shown by the white dashed line in~(e).}
\label{ARPES}
\end{figure*}

\begin{figure*}
\centering
\includegraphics[width=6.6in]{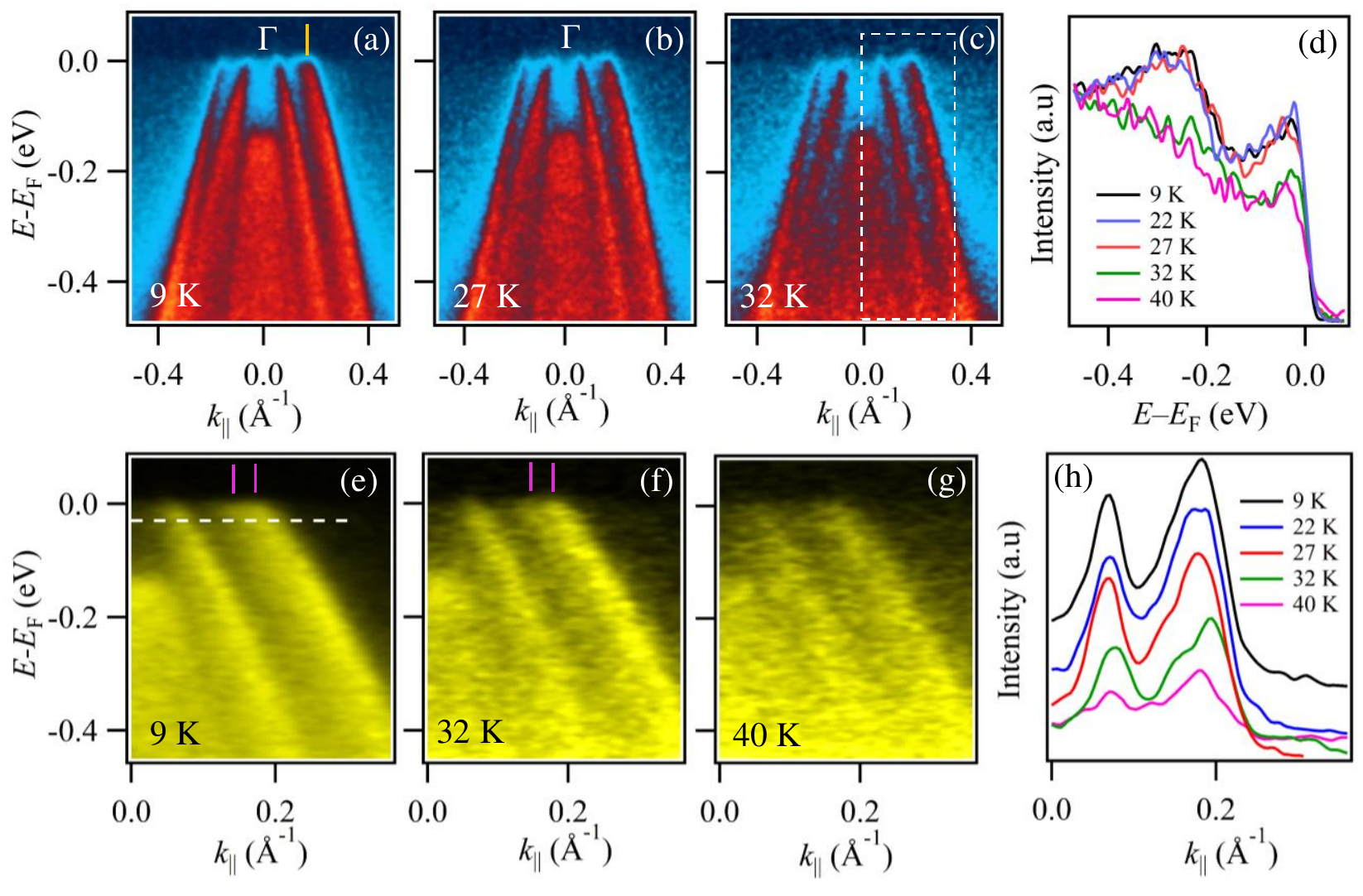}
\caption {Electronic structure of EuAl$_2$Ge$_2$ across the magnetic transition. (a)--(c) ARPES spectrum around ${\rm \Gamma}$ close to $E_{\rm F}$ along the cut shown by the dashed line in Fig.~\ref{ARPES}(e) for various temperatures 9~K, 27~K, and 32~K, respectively.  (d)~Temperature dependence of the energy distribution curves (EDCs) at the momentum indicated by a vertical line in (a). (e)--(g)~Zoomed view of the ARPES spectra within the region as indicated by a dashed rectangle in (c) for 9~K, 32~K, and 40~K, respectively. Arrows indicate the splitting of bands. (h)~Momentum distribution curves along the dashed line in~(e).}
\label{TD}
\end{figure*}

In order to understand the interplay of magnetism and electronic structure in \eag, ARPES measurements have been performed at different temperatures, with a particular emphasis on the temperature range bridging $T_{\rm N}$. The experimentally-observed electronic structure was also compared with the theoretical electronic structure by density-functional-theory (DFT)-based calculations.

Figure~\ref{ARPES}(a) shows the ARPES spectrum of \eag\ along the ${\rm {\Gamma}-{K}-{M}-{\Gamma}}$ path, measured in the AFM phase at $T = 9$~K\@. The spectrum shows two hole-like and one electron-like bands crossing the Fermi level at the ${\rm \Gamma}$ and M points of the Brillouin zone (BZ), respectively. These hole-like bands appear to cross at $-0.5$~eV along ${\rm {\Gamma}-{M}}$ [indicated by green arrow in (a)], but they are well separated along ${\rm {\Gamma}-{K}}$. For better visualization of the electron pocket, a  closer view is shown in the inset of Fig.~\ref{ARPES}(a). Extremely less-dispersive bands with high intensity are observed around $-1.5$~eV due to the localized Eu-$4f$ levels. Most of the experimental features are reasonably well-reproduced by DFT calculations, which considers the effect of spin-orbit-coupling (SOC) and a Hubbard $U = 5$~eV to account for the effect of strong localization of the half-filled Eu-$4f$ orbitals of \eag\ in its A-type AFM spin configuration as obtained from our neutron diffraction measurements [Fig.~\ref{ARPES}(b)].

In order to identify potential changes in the electronic structure associated with the magnetic transition, the AFM and paramagnetic (PM)  band structures are plotted together in Fig.~\ref{ARPES}(c). In the AFM phase, several new bands appear compared to the PM phase, due to the folding of electronic states originating from the doubling of the magnetic unit cell. For example, an electron-like band is observed in the AFM phase, just above $E_{\rm F}$ at the ${\rm \Gamma}$ point, whereas it is absent in the PM phase.  This electron-like band crosses two hole-like bands and causes various band anticrossings, as indicated by the arrows in the inset of Fig.~\ref{ARPES}(c).   Unfortunately, these states are inaccessible by photoemission spectroscopy as they appear above $E_{\rm F}$. However, potential changes in the electronic states between the PM and AFM phases are also expected below $E_{\rm F}$ as indicated by asterisk and triangle symbols with the dashed box, and arrow and that should be directly accessible by ARPES. Indeed we resolve those folded shallow bands in the AFM phase of \eag\ as indicated in Fig.~\ref{ARPES}(d), whereas no such states are observed in the PM phase. Generally, folded electronic states appear weaker in photoemission, regardless of whether they are due to magnetism or charge density waves~\cite{Ma2018, Ma2020, Schmitt2019, Brouet2004}.

Further, to map the dispersion of the electronic states in the $k_{x}$-$k_{y}$ plane, Fermi surface (FS) mapping was performed. Figures~\ref{ARPES}(e) and \ref{ARPES}(i) show the FS of \eag\ for the AFM and PM phases, respectively. In both cases, three Fermi pockets are observed, two at the center of the BZ, and one at the M~point. The circular and hexagonal Fermi pockets at the center of the BZ are formed by the inner and outer hole-like bands, respectively [Fig.~\ref{ARPES}(a)], and the elongated oval-shaped Fermi pocket at the M point is the electron pocket. This electron pocket is formed by the bottom of the conduction band that enters inside the Fermi level. The inner Fermi pocket is isotropic whereas the other two are very anisotropic that could produce the anisotropic magnetic properties as observed in our experiments.   All these FSs are well reproduced by theoretically-computed contours at $E_{\rm F} + 40$~meV [Figs.~\ref{ARPES}(g) and \ref{ARPES}(k)]. This energy shift was used to better match the shape and sizes of the experimental FS features, suggesting that the sample is slightly electron-doped.

The FS features and dispersion of electronic states suggest that \eag\ is metallic, both in the AFM and PM phases. Further, according to the band structure, folded bands between two consecutive BZs should connect the M~point at a deeper energy that cuts the folded bands at the M~point. Indeed, we observe this signature both in our ARPES and theoretical simulated constant-energy contours, as shown in Figs.~\ref{ARPES}(f) and \ref{ARPES}(h), respectively.  In the PM case, no such intensity is observed at the M~point due to the absence of band folding [Figs.~\ref{ARPES}(j) and \ref{ARPES}(i)]. Recently, magnetism-induced band folding and nontrivial band topology were reported in the Eu-based AFM system EuCd$_2$As$_2$~\cite{Ma2018, Ma2020}. As discussed above, our DFT calculations also predicted inverted band features in the AFM state of \eag\ near $E_{\rm F}$, which is typically observed in materials hosting nontrivial band topology. To correctly verify its nontrivial topological origin, we have calculated the $Z_2$ topological numbers using the Wilson loop (Wannier charge center) method~\cite{Mostofi2014} for the six time-reversal-invariant momentum planes. The obtained $Z_2$ topological numbers $v_0$;$(v_1v_2v_3)$ $=$ 1;(000) indicate the presence of nontrivial electronic states in this system.  Further theoretical studies are needed to determine the exact nature of the `topology' of the system.

To obtain more insight into the electronic structure change across the magnetic transition, we have performed high-resolution ARPES measurements close to $E_{\rm F}$ at various temperatures [Figs.~\ref{TD}(a)--\ref{TD}(c)]. While they exhibit very similar spectral features across the transition, the quasiparticle weight decreases significantly. This can be better visualized in their energy-distribution curves (EDCs) in Fig.~\ref{TD}(d). The temperatures at which the quasiparticle weight drops correlate well with magnetic transition temperatures. This drop in quasiparticle weight in the PM phase is most possibly related to the complex interplay between the orbital and spin degrees of freedom, caused by the change of coupling between magnetic moments and itinerant electrons across magnetic transitions.

Quasiparticle enhancement in magnetically-ordered states has been reported in other magnetic materials due to the decrease of spin fluctuations and changes in the scattering mechanism~\cite{Zhang2010, Jo2021}. Further zooming the ARPES spectra in momentum reveals that the individual hole-like bands actually split in two. The splitting is better resolved for the outer bands as indicated by vertical lines in Figs.~\ref{TD}(e) and \ref{TD}(f). The momentum distribution curves (MDCs) also show clear two-peak structures of the outer band. It is interesting to note that the band splitting survives above $T_{\rm N}$.

However, according to the theoretical calculations, all the bands in the AFM and PM phases are twofold degenerate, so no such band splitting is expected. Thus only two hole-like bands are expected to cross the $E_{\rm F}$ [Fig.~\ref{ARPES}(c)]. Generally, band splitting occurs when either time-reversal symmetry $T$ or parity $P$ symmetry is broken. Even though $T$ is broken in the AFM phase, the double degeneracy of the bands is protected by the combination of $P$, $T$, and translation ($L$) symmetries by one unit along the $c$axis~\cite{Ma2020}. The observation of band splitting in the PM phase is quite surprising as both the $T$ and $P$ symmetries should be preserved.

On the other hand, based on our magnetic measurements, the persistence of short-range FM correlations above $T_{\rm N}$ may cause the $T$ symmetry to break in the PM phase, leading to band splitting. In EuCd$_2$As$_2$, an analogous band splitting was reported~\cite{Ma2018}. The band splitting was explained as resulting from quasi-static and quasi-long-range FM fluctuations experienced by the itinerant electrons. In the AFM phase of \eag, the magnetic moments align ferromagnetically within a basal plane, which results from dominant in-plane FM exchange interactions. Since ARPES is a very surface-sensitive technique, these FM interactions may result in the band splitting in the magnetically-ordered state, as observed in Fig.~\ref{TD}(e).
\vspace{0.25in}
\section{\label{Sec:Conclu} Concluding Remarks}

We find that \eag\ is a metallic antiferromagnet with nontrivial electronic states in the AFM phase near $E_{\rm F}$. The compound exhibits A-type AFM order below $T_{\rm N} = 27.5(5)$~K with the Eu moments aligned in the $ab$~plane. The anisotropic magnetic properties exhibited by the system, associated with the Eu$^{2+}$ spins, indicate the presence of substantial magnetic dipole and magnetocrystalline anisotropy. The presence of in-plane magnetic anisotropy results in trigonal threefold AFM domain formation in $H = 0$. The moments in the domains exhibit a field-induced reorientation at \mbox{$H_{c1} \sim 2.5(1)$~kOe} to become perpendicular to the field direction for \mbox{$T < T_{\rm N}$}. The $ab$-plane and $c$-axis critical fields at $T = 2$~K are $H^{\rm c}_{ab} = 37.5(5)$~kOe and $H^{\rm c}_c = 52.5(5)$~kOe at which all moments polarized along the respective applied-field directions.

The presence of  dynamic short-range magnetic correlations within the $ab$~planes is evident above $T_{\rm N}$ from the zero-field heat capacity and resistivity studies.  A slight resistivity increase on cooling before loss of spin disorder scattering below $T_{\rm N}$ suggests magnetic correlations which are different from long-range AFM ordering.  Similarly, ARPES studies reveal band splitting even above $T_{\rm N}$, suggesting a possible breaking of the $T$ symmetry associated with the magnetic correlations above $T_{\rm N}$ which are therefore identified to be ferromagnetic in nature.  The ARPES results further reveal that \eag\ is metallic with a well-defined Fermi surface. The Fermi surface is formed by the two pockets at the zone center (${\rm \Gamma}$) and electron pockets at each M~point. The outer hole pocket and the electron pockets at M are very anisotropic. In addition to the various dispersive bands, extremely less-dispersive bands are observed around an energy $-1.5$~eV below the Fermi energy due to the localized Eu-$4f$ levels. Various folded bands are also observed in the AFM phase due to the doubling of the unit~cell. All these electronic states are modeled well by considering spin-orbit-coupling (SOC), $U$= 5 eV and the A-type $ab$-plane AFM configuration of the Eu magnetic moments.

\acknowledgments

The research at Ames National Laboratory was supported by the U.S. Department of Energy, Office of Basic Energy Sciences, Division of Materials Sciences and Engineering.  Ames National Laboratory is operated for the U.S. Department of Energy by Iowa State University under Contract No.~\mbox{DE-AC02-07CH11358.} The research at Brookhaven National Laboratory was supported by the U.S.\ Department of Energy, Office of Basic Energy Sciences, Contract \mbox{No.~DE-SC0012704.} This work was also supported in part by the Center for Spintronics Research Network, {\mbox{Tohoku} University.


\end{document}